\documentclass[a4paper,fleqn,usenatbib]{mnras}

\usepackage{newtxtext,newtxmath}

\usepackage{etoolbox}

\usepackage{xcolor}
\usepackage{amsmath}
\usepackage{graphicx}
\usepackage{psfrag}

\newcommand{\beq}{\begin{equation}}
\newcommand{\eeq}{\end{equation}}
\newcommand{\nn}{\nonumber}
\newcommand{\rmd}{\mathrm{d}}
\newcommand{\rmp}{\mathrm{p}}
\newcommand{\brac}[1]{\left({#1}\right)}
\newcommand{\pd}[2]{\frac{\partial{#1}}{\partial{#2}}}
\newcommand{\td}[2]{\frac{\rmd{#1}}{\rmd{#2}}}
\newcommand{\curl}{\nabla\times}
\renewcommand{\div}{\nabla\cdot}

\newcommand{\bB}{{\boldsymbol B}}

\newcommand{\be}{{\bf e}}

\newcommand{\sfM}{\mathsf{M}}
\newcommand{\sfL}{\mathsf{L}}
\newcommand{\sfT}{\mathsf{T}}
\renewcommand{\rmn}{\mathrm{n}}
\def\lesssim{\;\raise0.3ex\hbox{$<$\kern-0.75em\raise-1.1ex\hbox{$\sim$}}\;}
\def\gtrsim{\;\raise0.3ex\hbox{$>$\kern-0.75em\raise-1.1ex\hbox{$\sim$}}\;}
\def\mdens{{\rm g~cm^{-3}}}
\def\bdens{{\rm fm^{-3}}}

\def\nred{\overline{n}}

\title[Magnetic fields of PNSs]{Magnetic fields in late-stage proto-neutron stars}
\author[S. K. Lander et al.]{S. K. Lander${}^1$\thanks{samuel.lander@uea.ac.uk},
         P. Haensel${}^2$, B. Haskell${}^2$, J. L. Zdunik${}^2$ and M. Fortin${}^2$\\ \\
         ${}^1$School of Physics, University of East Anglia, Norwich, NR4 7TJ, U.K.\\
         ${}^2$Nicolaus Copernicus Astronomical Centre, Polish Academy of Sciences, Bartycka 18, 00-716 Warsaw, Poland}

\begin{document}

\pagerange{\pageref{firstpage}--\pageref{lastpage}} \pubyear{0000}
\maketitle

\label{firstpage}

\begin{abstract}
We explore the thermal and magnetic-field structure of a late-stage proto-neutron
star. We find the dominant contribution to the
entropy in different regions of the star, from which we build a
simplified equation of state for the hot neutron star. With this, we
numerically solve the stellar equilibrium equations to find a range of
models, including magnetic fields and rotation up to Keplerian
velocity. We approximate the equation of state as a barotrope,
  and discuss the validity of this assumption. For fixed magnetic-field strength, the
induced ellipticity increases with temperature; we give quantitative
formulae for this. The Keplerian velocity is 
considerably lower for hotter stars, which may set a de-facto maximum
rotation rate for non-recycled NSs well below 1 kHz.
Magnetic fields stronger than around $10^{14}$ G have
  qualitatively similar equilibrium states in both hot and cold
  neutron stars, with large-scale simple structure and the
poloidal field component dominating over the toroidal one; we argue
this result may be universal. We show that truncating magnetic-field
solutions at low multipoles leads to serious inaccuracies, especially
for models with rapid rotation or a strong toroidal-field component.
\end{abstract}


\maketitle

\section{Introduction}

In the first phase of its life, a highly-magnetised neutron star (NS) has the
potential to radiate a huge amount of energy, through both
electromagnetic and gravitational waves. These signals are of great
interest, containing information that could allow us to constrain
processes involving elementary constituents of matter
under extreme astrophysical conditions, the nuclear physics of hot dense matter, the fluid
dynamics of the newborn star, and the
dynamo processes driving magnetic-field amplification in extremely
highly-conducting media.

With their astrophysical importance and complexity, supernovae and
proto-neutron stars have long been studied through numerical 
evolutions (see e.g. \citet{colgate,BL86,janka}), and their
hydrodynamics and microphysics -- among other aspects -- remain topics
of active study. By contrast, the magnetic field
of the newborn NS has received relatively little attention, especially
given that this phase is likely to be the most dramatic of the field's life. It is likely
that some remnant field of the progenitor star is amplified and
rearranged during this phase \citep{TD93}, but we lack any quantitative
understanding of this process.

For a cooling, mature NS we have a better -- though still incomplete
-- understanding of its magnetic field. In particular, a reasonably
complete picture of magnetic-field evolution within the star's solid
crust has emerged after sustained attention; see
\citet{ponsvig_review} for a recent review. Core
evolution is far less certain, though may be too slow to be of
relevance to many problems. Complementary to these evolutions are a number of studies of
possible equilibrium states of a magnetised NS, solving for the global
field but without accounting for the evolutionary history frozen
into the crust; for a brief but representative selection of these
models see, e.g., \citet{bocquet,KKY,ciolfi10,L14,GLA,pili15}.

In comparison with the attention shown to the star's birth and
maturity, the late proto-NS phase (covering a period 
from  some ten seconds to  roughly  a few  minutes after birth) is terra incognita,
especially for the star's magnetic field. It
may, however, be a very important stage in the star's evolution: one
where the physics driving the star's birth phase will have
ceased, but thermal effects will still be important. Magnetic-field
rearrangement during this early era, rather than any dynamo mechanism, may
be what sets the basic long-term structure of the mature star's
field. The resultant field configurations would also be the logical
initial condition for field-evolution studies. In this paper we aim
to explore the late proto-NS phase in more detail, looking at the main
contributions to the star's thermal structure and finding equilibrium
states for a magnetised NS at high temperature.

\subsection{Supernova and aftermath}

The life of a massive star culminates in the gravitational collapse of
its core. If the star's mass is more than a few tens of times that of
the Sun, the collapse
continues unabated until a black hole is formed. Otherwise, the
compression of matter is
brought to a halt by the high stiffness (incompressibility) of uniform nuclear matter, causing a
bounce. This occurs on a surface enclosing the denser half of the mass of the future
proto NS and sends a shock wave through the envelope,
heating it strongly and lifting infalling stellar
matter, and thus separating a hot and dense central object from
the pre-supernova star doomed to explosion \citep{woosley02}. A proto NS is born.  Its
initial  internal temperature, $T$,   and  entropy per baryon, $s_b$,
are very non-uniform, with maxima reached in the shocked half of the
mass.

Initially, a proto NS  is opaque to neutrinos, and the total
electron lepton number per baryon $Y_{L_{\textrm e}}$ (a sum of electron $Y_{\textrm e}$
and electron neutrino $Y_{\nu_{\textrm e}}$ contributions) is
$\approx 0.35$, only slightly  smaller than in the presupernova
iron-nickel core.  During the next  seconds the proto-NS deleptonizes via
$\nu_{\textrm e}$  diffusion driven  by the  $Y_{\nu_{\textrm e}}$  gradient.  The
$\nu_{\textrm e}$  thermalize, losing their degeneracy, and leave the star through
the \emph{neutrinosphere} (the surface at which matter becomes
neutrino-transparent). The transport of lepton number and energy by
diffusion is accelerated by convective flows. Diffusion of $\nu_{\textrm e}$
outwards  is associated  with heating of the matter  by the  $\nu_{\textrm e}$
downscattering.  After $\sim 10$ s  deleptonization has been completed,
gradients of $T$  and $s_b$ smoothed, and convective stability reached.
Neutrino-antineutrino pairs of all three flavours still transport heat
via diffusion towards the neutrinosphere, and are radiated
there \citep{prakash97}. The proto NS enters its late stage, the subject of the present
study.

During the next minute or so, with a composition not very different
from that of a mature NS, the proto NS is still  hot,  $\sim 5\times
10^{10}~$K in the core, with an envelope composed  of a  plasma of  nuclei,
neutrons, and electrons, and density  above $10^{11}~\mdens$. The
envelope is neutrino-opaque,  and layers above it 
contain  the flavour-dependent, rather thick,  neutrinospheres. The
envelope is liquid, even in its deepest layers close to the core. Its
temperature is decreasing outwards. As we assume slow neutrino cooling
(no direct Urca in the core), at this late  proto NS stage  both $T$
and $s_b$  are slowly varying within the core, decreasing more
rapidly towards the neutrinospheres.  In the present paper we assume
that there is no plasma fallback after a successful shock take-off. 

The pressure in a mature NS core  is due to nuclear forces and to a
lesser extent, to the degeneracy of the neutrons \citep{cam59}.  In the inner
envelope, where nuclei are immersed in a neutron gas,  the pressure is
supplied by the degeneracy of the neutron gas, with the contribution from
nuclear forces in  dripped  neutron gas rapidly increasing close to
the core. This is in contrast to normal, non-degenerate stars, where
pressure is thermal in nature; these stars are hot, powered by fusion
processes. The  proto-NS  phase has the distinguishing feature
that both neutron-degeneracy and thermal pressure play a role in
determining the stellar structure; with neutrinos flooding out of the newborn star once
it becomes neutrino-transparent, this phase is over within a few minutes
 \citep{BL86,pons99}. However,
this brief period of time -- which has not been previously explored in the
context of magnetic-field modelling -- is a crucial one to understand.
It constitutes a missing link between work on the dynamic evolution
and generation of magnetic fields in proto-NSs, as described next, and the far
slower, secular evolution in mature neutron stars.

The minute following a NS's birth is crucial for the star's magnetic field. The
magnetic field of the progenitor star's degenerate core will be 
amplified by compression to nuclear densities during stellar core 
collapse, but this alone is
unlikely to explain the field strengths of NSs, especially magnetars,
where the external field is around $10^{15}$ G and the internal field
perhaps an order of magnitude stronger. Instead, dynamo processes act
to amplify and rearrange the field; these could involve some
combination of differential rotation, convection and the
magneto-rotational instability (for a review of this topic, see \citet{spruit09}). These processes are likely to cease at
a very early phase, with the dynamo saturating and becoming inhibited,
magnetic coupling flattening the rotation
profile so the star's rotation becomes uniform, and turbulent
convection ceasing as the stellar matter becomes neutrino-transparent.

\subsection{The early quasi-equilibrium}
\label{quasi_eq}

Whatever magnetic field has been created in the birth phase will afterwards
start to rearrange, in an attempt to attain an approximate equilibrium with the
fluid star. It is perhaps possible that it fails to settle in this
  way and instead reaches some kind of `average' steady state, where the star
still exhibits short-term dynamics but average values of energy quantities are roughly
constant \citep{sur}, though it is likely that over longer timescales
this would dissipate considerable amounts of energy. Here we will
assume that the magnetised star does indeed reach a true equilibrium
-- one that is also dynamically
stable, and so a natural endpoint for the rearrangement.

To establish when the star can be treated as in approximate
equilibrium, we first need to know how quickly the magnetic field can
adapt to its host fluid. Assuming, for the time being, a non-rotating
star,  the field is able to rearrange locally
over the time $\tau_A$ taken for an Alfv\'en wave to cross the region
concerned. Defining $l_B$ as the distance crossed and $v_A$ as the
Alfv\'en speed, we have
\begin{align}
\tau_A\sim\frac{l_B}{v_A} \sim &\frac{l_B\sqrt{4\pi\rho}}{B}\nn\\
\approx & 1.1\brac{\frac{B}{10^{14}\textrm{ G}}}^{-1}
                \brac{\frac{\rho}{10^{15}\textrm{ g cm}^{-3}}}^{1/2}
                \brac{\frac{l_B}{10\textrm{ km}}}\textrm{ s}
 \label{tau_B}
\end{align}
where $B$ is the magnetic field strength, $\rho$ the rest-mass
density, and where we have normalised $l_B$ to 10 km to get a timescale for
global magnetic-field rearrangement. Implicit in the above estimate is
that the star is entirely fluid; we confirm this in section
\ref{env_state}, with quantitative calculations of the state of matter of the
proto-NS.

Note that this estimate may not be reliable for the very rapidly
  rotating models we consider later in this paper; it is known, for
  example, that in the presence of rotation Alfv\'en oscillation modes
  are replaced by magneto-inertial modes (see, e.g., \citet{lin_ogilvie} and
  references therein). These modes tend to be of higher
  frequency than their non-rotating counterparts \citep{LJP10,LJ11}, leading to a shorter
  $\tau_A$ and thus suggesting that the magnetic field may be able to
  re-equilibrate to the fluid more quickly at increasing rotation
  rate. The opposite conclusion was, however, reached by
  \citet{braith_cant} from timescale arguments, although these authors
  made the simplication of not explicitly considering the centrifugal
  distortion to the star. Fortunately this uncertainty does not have
  any serious impact on our main results.

Now, if the Alfv\'en timescale $\tau_A$ is short compared
with the cooling timescale of the star, the magnetic field should always have time to
readjust to the new thermal state of the star, and therefore may be
thought of as proceeding through a sequence of
quasi-equilibria. In this case, an individual equilibrium model
  may be thought of as a snapshot of this process. We defer the more
  technical discussion of the role of chemical equilibrium to section \ref{eqm_EOS}.

We can make a rough estimate of the cooling timescale for a proto-NS
from visual inspection of the plots of \citet{BL86} or similar work;
it is of order $10$ s. Therefore, for large-scale magnetic fields
stronger than roughly $10^{14}$ G, the equilibrium approximation is reasonable even during this
early phase. For weaker magnetic fields, it is possible that the field
will spend this phase out-of-equilibrium with the fluid, retaining
vestiges of the (presumably) complex magnetic-field structure produced
by the birth. However, it is quite
plausible that $10^{14}$ G does represent a typical birth magnetic-field
strength, with the typical surface field strength decaying to
pulsar-type values by the time we observe them. In any case, we
consider here that class of late proto-NSs for whom a
quasi-equilibrium approximation is reasonable.

\subsection{Plan of the paper}

This paper is arranged as follows. In section \ref{therm_EOS_sec} we begin with a description of the
equation of state and thermal physics of a hot neutron star, and describe
more precisely the meaning of the `late' proto-neutron star phase. We devise a simplified
model of the thermal physics of the star, retaining the leading-order
contributions in each region. In section \ref{eqm_equations} we discuss the general equilibrium
equations and the equation of state, in particular the possible
  presence of buoyancy forces, and in section \ref{Ptherm_sec} we describe our prescription for
the thermal pressure. Section \ref{numerics} formulates the problem in
a way we can solve numerically, and we give details of this solution
method. Our results are presented in section \ref{results}, and we
discuss their implications in section \ref{discussion}.

\section{Thermal structure of a late proto-neutron star}
\label{therm_EOS_sec}

\subsection{Equation of state of a late proto-neutron star}
\label{therm_EOS}

\subsubsection{Equation of state of the core}
\label{sect:therm.core}

The core consists of a uniform plasma  of mainly neutrons, with a
small  admixture of protons, electrons and muons.  Thermodynamical
quantities, such as  internal energy per unit volume $U$, pressure
$P$,....  are  split   into  a  $T=0$  (cold) part, $U_0$,
$P_0$,\ldots  and a thermal contribution depending  on $T$ and
vanishing in the $T=0$ limit, e.g., 
$U_{\rm th},P_{\rm th},\ldots$. For the  $T=0$ equation of state (EOS)  we choose an
approximation of the SLy EOS \citep{DH01} by a piecewise
polytrope. Then we get the  $T = 0$  values of the   baryon chemical
potential  from $\mu_0 = m_{\rm u}c^2 + (U_0 + P_0)/n$  and  the matter
density, including  rest energy of nucleons,  $\rho_0= m_{\rm u}n +
U_0/c^2$, where  $n$ is the  
baryon number density, and $m_{\rm u}$ is the atomic mass unit.   
The  thermal components of the core EOS  are approximated  by  those
of an ideal, nonrelativistic,  strongly degenerate Fermi gas of
neutrons, with number density $n$.
The Fermi momentum and Fermi energy of a degenerate gas of free neutrons, $p_{\rm Fn}$, 
are related to $n$ by $p_{\rm Fn}=\hbar(3\pi^2 n)^{1/3}$, 
$\varepsilon_{\rm Fn}=\hbar^2(3\pi^2 n)^{2/3}/(2m_{\rm u})$.
Because of the 
supranuclear densities prevailing in the core, it  is convenient to
express  $n$  in the units  of normal nuclear density $n_0 =
0.16~\bdens$. We therefore define $\overline{n} = n/n_0$. The core
edge is at about $\overline{n} = 0.5$.  The  Fermi energy and Fermi
temperature of neutrons are in our approximation
\begin{eqnarray}
\varepsilon_{{\rm Fn}}=58.44~ {\overline{n}}^{2/3}~{\rm MeV}~,~~ 
T_{\rm F}=\varepsilon_{{\rm Fn}}/k_{\rm B}~,\cr\cr 
~~T/T_{{\rm Fn}}=1.47\times 10^{-2}~T_{10}/\overline{n}^{2/3}~,
\end{eqnarray}
where  $T_{10}=T/10^{10}~{\rm K}$.
To find the thermal contributions to $U$,$P$, entropy density $S$, and 
neutron chemical potential $\mu$, we take those derived for a degenerate free nonrelativistic electron 
gas (\S 58 of \citet{LL-SP1}) and replace the electron mass by the neutron
one. Then, neglecting powers of $T/T_{{\rm Fn}}$ higher
than two, we get:
\begin{eqnarray}
U_{\rm th}={1\over 4}\pi^2 n \varepsilon_{\rm Fn} (T/T_{\rm Fn})^2~,~~ 
P_{\rm th}={1\over 6}\pi^2 n \varepsilon_{\rm Fn} (T/T_{\rm Fn})^2~,\cr\cr
S_{\rm th}=S={1\over 2}\pi^2 n k_{\rm B}{T/T_{\rm Fn}}~,~~ 
\mu_{\rm th}=-{1\over 12}\pi^2\varepsilon_{\rm Fn} (T/T_{\rm Fn})^2~.
\label{eq:core.n-therm}
\end{eqnarray}

Note that   $S_0=0$ and  $\mu_0=\varepsilon_{{\rm Fn}}$.

\subsubsection{Equation of state  and composition of the  envelope}
\label{sect:eos-compos-envel}
At $T=0$  the envelope  is a solid crust of nuclei localized in the
crystal lattice sites. Under the
conditions prevailing  in the late stage of a proto-NS, this `crust'
will in fact be a liquid envelope.  In contrast to the  core, the  envelope   is  a
nonuniform  form of dense matter.  It consists of 
nuclei, which  for densities larger than the neutron drip density,
$n_{\rm nd}$, (i.e., in the inner envelope)  are immersed in a gas of 
unbound neutrons.   At a given baryon density $n$ the envelope  is treated  as a  plasma 
of  one type of ions (nuclei),  possibly  immersed in an neutron gas,
all permeated by  a quasi-uniform electron gas. Within the envelope, we  will use an
approximate  relation between  $n$  and the matter density
$\rho\simeq n m_{\rm u}$.
 We will use  the SLy model  of the $T=0$ crust \citep{DH01}, for
 consistency with our core prescription.
As in the case of the uniform liquid core, the  EOS of the envelope is split into a 
$T=0$ (cold)  part,  $U_0(\rho), P_0(\rho)$ and a thermal  one,
$U_{\rm th}(\rho,T),P_{\rm th}(\rho,T)$   vanishing in the limit of
$T=0$.   We introduce a set 
of parameters   characterizing  locally  this  layer of a late-stage  proto-NS.  These parameters are 
functions of the density $\rho$. The number density of ions  
(nuclei) is  $n_{\rm  i}$.  The number of nucleons  and  number of
protons in an ion are $A$ and $Z$, respectively. We define an ion
sphere   of radius  $a_{\rm i}$ such that  its volume
$\frac{4}{3}\pi a_{\rm i}^3$  is equal to the volume per ion
$1/n_{\rm i}$.
The ion sphere  contains a single ion at its centre and $Z$ electrons that  neutralize  the
ion charge $Ze$.   In the  inner envelope  a fraction of neutrons is
unbound, and therefore  the number of nucleons in an  ion sphere  is
$A^\prime>A$.   One must therefore specify  the fraction of unbound neutrons  in the total number
of nucleons, $X_{\rm n}$. In the outer envelope $X_{\rm n}=0$.  The fraction of
volume occupied by  nuclei will be denoted by $u$. 

In what follows we derive the thermal part  of  the EOS of the envelope, to be added to the 
dominant $T=0$ part. Our notation follows 
Ch.2,3 of \citet{NSBook}. We consider ions, unbound neutrons, electrons
and their contributions to the  thermodynamic quantities  in the $T -
\rho$ plane. We do not include thermal effects on the
composition, which for our range of $T$ is reasonable for  ${\rm log}(\rho/\mdens) > 10$.

We start with  the simplest component of the  envelope: the electrons.   
Already  for  ${\rm log}(\rho/\mdens) > 8$ electrons form a (nearly)
uniform ultrarelativistic  quasifree  Fermi gas with Fermi energy
\begin{equation}
\varepsilon_{{\rm Fe}} = p_{{\rm Fe}}c = 33.14~(n_{\textrm e}/10^{-3}n_0)^{1/3}~{\rm MeV}~.
\end{equation}
At neutron drip $\nred = \nred_{\rm nd}\approx 10^{-3}$ and $n_{\textrm e}\approx 0.3n_{\rm nd}$ so that 
$\varepsilon_{{\rm Fe}}^{\rm nd}\approx 22~{\rm MeV}$.

In our case, with ${\rm log}(\rho/\mdens) > 10$,   the electrons  are strongly degenerate, 
$T\ll T_{{\rm Fe}}=\varepsilon_{{\rm Fe}}/k_{\rm B}$.  Keeping  only  the leading 
terms of an expansion in $T/T_{{\rm Fe}}$, we  get the following
formulae  for the thermal contribution of the electrons (see \S 61 of \citet{LL-SP1}):

\begin{eqnarray}
U_{\rm th}={1\over 2}\pi^2 n_{\textrm e} \varepsilon_{\rm F} 
\left({T/T_{{\rm Fe}}}\right)^2~, ~
~~P_{\rm th}=
{1\over 6}\pi^2 n_{\textrm e} \varepsilon_{{\rm Fe}}
\left({T/T_{{\rm Fe}}}\right)^2~,\cr\cr
S_{\textrm e}=\pi^2 n_{\textrm e} k_{\rm B} ~{T/ T_{{\rm Fe}}}~,~~~
\mu_{\rm th}= -{1\over 3}\pi^2\varepsilon_{{\rm Fe}}\left({T/T_{{\rm Fe}}}\right)^2~,\cr\cr~~
T_{{\rm Fe}}=38.46\times 10^{10}(n_{\textrm e}/10^{-3}n_0)^{1/3}~{\rm K}~.
\end{eqnarray}

Next we turn to the ion component of the envelope, for which we need
to define and calculate various parameters. Firstly, the  number density of ions is  expressed as  
$n_{\rm i} = \rho/(A^\prime m_{\rm u})$, and average charge neutrality
implies  $n_{\textrm e} = n_{\rm i}Z$.  From these, we can now express
the ion sphere radius $a_{\rm i}$ to plasma parameters in two ways:
\begin{equation}
{4\over 3}\pi a_{\rm i}^3 n = n_{\rm i} A^\prime/n = n_{\rm i} Z/n_{\rm e}~~.
\end{equation}

We can understand the state of matter in the envelope through the
dimensionless \emph{Coulomb coupling parameter} for ions,
which measures the relative strength of the Coulomb interaction of
ions compared to the energy of their thermal motion:
\begin{equation}
\Gamma_{\rm i}={Z^2 e^2\over a_{\rm i}k_{\rm B}T}~.
\end{equation}
The strength of correlations between ions in the envelope and their contribution 
to the ion thermodynamical  quantities can be expressed in terms
of $\Gamma_{\rm i}$. The numerical value  of  $\Gamma_{\rm i}$
for a plasma can be 
readily obtained by passing to dimensionless variables:
\begin{equation}
\Gamma_{\rm i}={7.42\over T_9}\left({\rho_{10}\over A^\prime/100}\right)^{1/3} 
\left({Z\over 40}\right)^2 ~,
\label{eq.Gamma_i.num}
\end{equation}
where   $\rho_{10}=\rho/10^{10}\;\mdens$  and  $T_9=T/10^9\;{\rm K}$.

Using $\Gamma_{\rm i}$, we distinguish three main physical  regimes of the 
plasma  in the  density-temperature plane.
If  $\Gamma_{\rm i}\ll 1$ then Coulomb correlations between ions are 
unimportant,  and the thermal state of ions  is well approximated  by a Boltzmann
gas model.   Coulomb correlations become important when  $\Gamma_{\rm i}\simeq 1$, 
and grow stronger and stronger with increasing $\Gamma_{\rm i}$. For a given $\rho$, $\Gamma_{\rm i}=1$ is 
reached at a characteristic temperature $T_l$ that may be found
  by rearranging Eq. \eqref{eq.Gamma_i.num}:
\begin{equation}
T_l= 9.504\times 10^{10}\left( {Z\over 30}\right)^2 
\left({\rho_{10}\over 
{A^\prime/100}}\right)^{1/3}~{\rm K}~.
\end{equation}
So at a given $\rho$, the ions behave as a nearly-ideal Boltzmann gas of nuclei if 
$T\gg T_l$. Then for smaller values of $T$ within $T_m<T\lesssim 
T_l$ (where $T_m$ is the melting temperature of an ion
crystal)  correlations are important and we are dealing with a
strongly-coupled Coulomb liquid of ions. Finally, at an even lower $T=T_m$ the Coulomb
liquid of ions crystallizes (solidifies) via a first order phase transition, 
with very small  latent heat. Numerical  simulations predict that to a
very good approximation the free energy of the ion liquid (with quantum
contributions negligible) is a function of $\Gamma_{\rm i}$ only. Crystallization occurs at 
$\Gamma_{\rm i}=175$ \citep{PotCha2000}, which leads to (again using Eq.\;\eqref{eq.Gamma_i.num}):
\begin{equation}
T_m=5.43\times 10^8 \left( {Z\over 30}\right)^2 
\left({\rho_{10}\over {A^\prime/100}}\right)^{1/3}~{\rm  K}~.
\end{equation}
There is an additional plasma  parameter that allows one to determine  
the relative  importance of quantum effects in the  thermal
properties of the ion liquid. This is the plasma frequency for the ions
$\omega_{\rm pi}$, corresponding to the frequency of vibrations  generated by  shifting an ion
from the equilibrium position, $\omega_{{\rm pi}}=(4\pi e^2 n_{\rm i}Z^2/M_{\rm i})^{1/2}$, 
where $M_{\rm i}=A m_{\rm u}$ is the ion (nucleus) mass.
After dividing $\hbar \omega_{\rm pi}$ by $k_{\rm B}$ we 
get a characteristic temperature $T_{\rm pi}$,
\begin{equation}
T_{\rm pi}={\hbar \omega_{\rm pi}\over k_{\rm B}}=
4.95\times  10^7 \left( {({Z/40})^2 \over {A/100}}\right)^{1/2} 
\left({\rho_{10}\over {A^\prime/100}}\right)^{1/2}~{\rm  K}~.
\end{equation}
For $T\gg T_{\rm pi}$ a classical treatment of the ion motion is
valid -- and this is 
the case for the envelopes under consideration here. 

Another  important   ionic parameter  is  the thermal de Broglie wavelength, 
appearing in the  formula  for the chemical potential and the entropy
of the Boltzmann gas of massive particles (\S 45 of \citet{LL-SP1}). It is given by:
\begin{equation}
\lambda_{\rm i}=\left({2\pi\hbar^2}\over{M_{\rm i}k_{\rm B}T}\right)^{1/2}.
\end{equation}
This formula is strictly valid in the outer envelope. More
  generally,  in the presence of unbound neutrons, the number density
  of ions  is  related to the mass density of the plasma by
\begin{equation}
n_{\rm i}=\rho/(A^\prime m_{\rm u})= 0.597~\rho_{11}/({A^\prime/100})
\times 10^{33}~{\rm cm^{-3}}~. 
\end{equation}

What matters for  the  chemical  potential of ions, $\mu_{\rm i}$, and the entropy density, 
$S_{\rm i}$, is a dimensionless parameter $n_{\rm i}\lambda_{\rm i}^3$.  It  plays a double
role. First, it enters the formulae for $\mu_{\rm i}$ and  $S_{\rm i}$. 
For the Boltzmann gas of ions we have
\begin{equation}
\mu_{\rm i}=k_{\rm B}T~{\rm ln}(n_{\rm i}\lambda_{\rm i}^3)~,~~
S_{\rm i}={5\over 2}k_{\rm B}n_{\rm i}-k_{\rm B}n_{\rm i}
~{\rm ln}(n_{\rm i}\lambda_{\rm i}^3)~.
\label{eq:mu.S.ions}
\end{equation}
Second, when $n_{\rm i}\lambda_{\rm i}^3\lll 1$~, then $\mu_{\rm i}$ is large 
and negative, and this tells us that Boltzmann statistics is valid. The  ideal
Boltzmann gas formulae for  the ion contributions to  $C_V$ and $P_{\rm th}$ are then  valid:
\begin{equation}
C_{V{\rm i}}={3\over 2}k_{\rm B}n_{\rm i}~, ~P^{\rm i}_{\rm th}=n_{\rm i}k_{\rm B}T~. 
\end{equation}
At first glance, it may seem that  the  contribution  of the Coulomb
interaction  (correlations) between ions, and between ions and
electron gas,  has to be added to the  
ideal Boltzmann gas quantities  for the ions.   As we already  mentioned,  these 
Coulomb interaction contributions   can be expressed in terms of the 
Coulomb coupling parameter  $\Gamma_{\rm i}$. 
In our case  $\Gamma_{\rm i}\gg 1$  (i.e. a strongly coupled  Coulomb liquid of ions)  
and the leading   Coulomb contribution, denoted  as  $U_{\rm ii}$,  is
\citep{PotCha2000}
\begin{equation}
 U_{\rm ii}=k_{\rm B}T n_{\rm i}\Gamma_{\rm i}^{3/2}=n_{\rm i}A_1 Z^2e^2/a_{\rm i}~,~~~
 A_1=-0.9070~.
 \end{equation}
 So, at this approximation there is no  $T$ dependence  of the Coulomb
 contribution and therefore there is  no need to modify our formulae
 for $U_{\rm th}$. Actually, $U_{\rm ii}$ 
has already been included in our $U_{0}$, $P_{0}$ as the so called lattice term.

Our last  component  of the thermal part of the EOS of the envelope  comes from unbound 
neutrons. We neglect  contributions from evaporated protons and alpha particles; their
populations are small compared to that of  the unbound neutrons at the densities and $T$ 
relevant  for the late-stage proto-NS envelope.   In the inner
envelope we add contributions from the neutron gas of density
$n_{\rm ng}$ (this is the density measured in the space outside nuclei). This  gas is degenerate  except for a layer 
close to the   neutron 
drip point,  $n\approx n_{\rm nd}$.  The  contribution from this  thin
non-degenerate layer will be neglected.  Apart from this neglected
layer, unbound neutrons outside nuclei  form  a degenerate
non-relativistic Fermi gas,  filling  the available volume  outside
nuclei, with microscopic number density   
\begin{equation}
n_{\rm ng}={{X_{\rm n} {n}}/({1-u})}~,~
\end{equation}
where $X_{\rm n}$ is the unbound neutron fraction relative to all nucleons,  and  $u$ is the 
volume fraction occupied by nuclei. We approximate the Fermi energy and Fermi temperature
of the neutron gas by the free Fermi gas values (Eq.\;\eqref{eq:core.n-therm})
\begin{eqnarray}
\varepsilon_{{\rm F}}^{\rm ng}=58.44~(\overline{n}_{\rm ng})^{2/3}~{\rm MeV},~~
T_{{\rm F}}^{\rm ng}=\varepsilon_{{\rm F}}^{\rm ng}/k_{\rm B}~,\cr\cr
~~ T/T_{\rm F}^{\rm ng}=1.47\times 10^{-2}~T_{10}/(\overline{n}_{\rm ng})^{2/3}~~.
\end{eqnarray}

Keeping only leading terms with respect to a small degeneracy parameter 
$T/T_{\rm F}^{\rm ng}$, we obtain, using Eq.\;\eqref{eq:core.n-therm},  approximate expressions for 
 $U_{\rm th}^{\rm ng}$, $P_{\rm th}^{\rm ng}$, $S_{\rm ng}$, and
 $C_V^{\rm ng}$, per unit volume  of the dripped neutron gas (i.e.,
 with the volume of nuclei being excluded):
\begin{eqnarray}
U_{\rm th}^{\rm ng}={1\over 4}\pi^2 n_{\rm ng}~\varepsilon_{\rm F}^{\rm ng}
\left(T/T_{\rm F}^{\rm ng}\right)^2~,\cr\cr ~~
P_{\rm th}^{\rm ng}={1\over 6}\pi^2 n_{\rm ng}~\varepsilon_{\rm F}^{\rm ng}
\left(T/T_{\rm F}^{\rm ng}\right)^2~,\cr\cr
S_{\rm ng}=C_V^{\rm ng}=
{1\over 2}\pi^2 n_{\rm ng}~k_B
T/T_{\rm F}^{\rm ng}~,\cr\cr~~
\mu_{\rm th}^{\rm ng}=-{1\over 12}\pi^2 ~\varepsilon_{\rm F}^{\rm ng}
\left(T/T_{\rm F}^{\rm ng}\right)^2~.
\label{eq:n-dripped.therm}
\end{eqnarray}
The contribution to the total (macroscopic)  $U_{\rm th}$, $P_{\rm th}$, $S$, $C_V$ can be obtained 
by multiplying   the quantities given  in
Eq.(\ref{eq:n-dripped.therm})  by a factor  $(1-u)$.  Note that even
at the bottom of the inner crust $u<0.3$, so later we will neglect $u$
corrections to simplify our calculations 
(nucleon effective mass corrections, which are also neglected, are of
a similar size to the $u$-ones). 

\subsection{Relative importance of different entropy contributions}

\begin{figure*}
\begin{center}
\begin{minipage}[c]{0.9\linewidth}
  \includegraphics[width=0.47\linewidth]{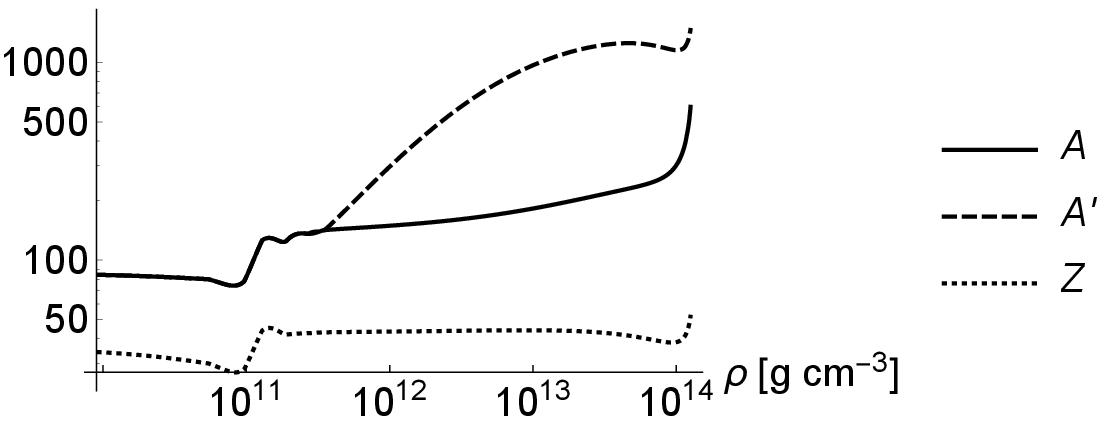}
  \hfill
\includegraphics[width=0.47\linewidth]{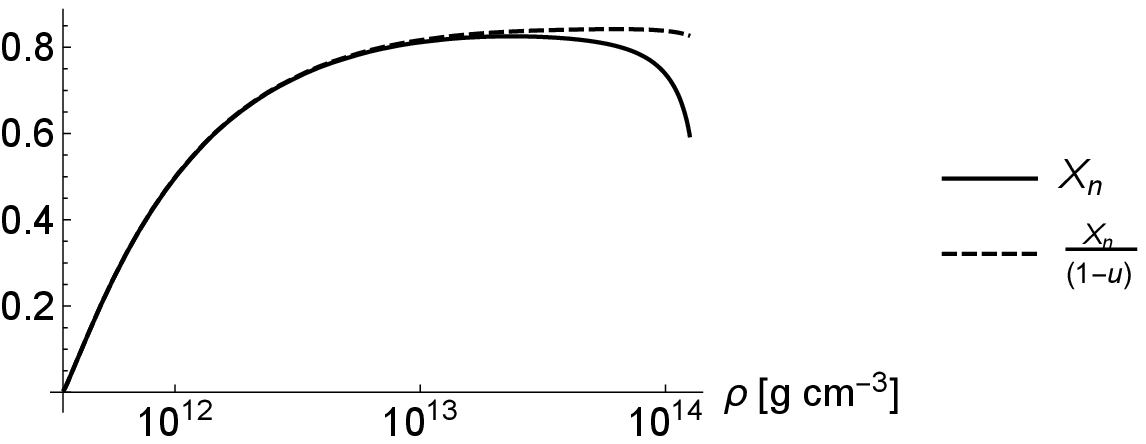}
\caption{\label{env_quant}
  Interpolations to various envelope quantities as a function of
  density. Note that $X_{\rm n}\to 0$ at $\rho_{\rm nd}$, but
  $X_{\rm n}\to\!\!\!\!\!\! /\ \ 1$
at $\rho_{\rm cc}$.}
\end{minipage}
\end{center}
\end{figure*}

To approximate the thermal structure of a hot, late-stage proto-NS, we need
to ascertain the relative importance of different components in the
various regions of the star. We divide the star into three regions:\\
(i) the core, $\rho\geq\rho_{\rm cc}$;\\
(ii) the inner envelope, $\rho_{\rm nd}<\rho<\rho_{\rm cc}$;\\
(iii) the outer envelope, $0<\rho\leq\rho_{\rm nd}$;\\
where
\begin{align}
  \rho_{\rm nd}&=3.5\times 10^{11}\textrm{g cm}^{-3},\nn\\
  \rho_{\rm cc}&=1.4\times 10^{14}\textrm{g cm}^{-3}.\nn
\end{align}
We give these quantities the subscripts nd and cc, since they
correspond to the neutron-drip point and crust-core density for a
mature neutron star, although the terms should not be taken too
literally here; the stellar structure shortly after
birth is complex, the transition densities less clearly-defined, and
the crust has not yet begun forming.

As discussed in section \ref{sect:therm.core}, it is clear that in the
core the degenerate baryons provide the
dominant contribution to the thermal structure \citep{BL86,pons99}, 
and since the   majority of these are neutrons, it is a safe
first approximation to model the core entropy as being due to
degenerate neutrons alone.  At the temperatures 
under consideration they are in a non-superfluid (normal) state. 

Our model will be far simpler if the entropy contribution from one
particular species is dominant in each of the different envelope
regions too. This is not
guaranteed, however, so we now proceed to evaluate these contributions
to check.

\subsubsection{Interpolated envelope equation of state}

To calculate the thermal contributions to the envelope, we need
various equation-of-state quantities: $A,Z,A',X_\rmn$ and $u$ as a function of
$\rho$. To construct smooth functions for these dependences, we use
inbuilt fitting routines from the software package Mathematica to make 
interpolations of tabulated equation-of-state data from \citet{DH01}
for the inner envelope, and \citet{haen-pich} for the outer
envelope. The fitting functions to the different envelope quantities
are plotted in figure \ref{env_quant}. Note that for our model of the
core the quantity $X_\rmn/(1-u)$ is effectively equal to unity,
whereas at the inner edge of the envelope it is roughly $0.8$.

\subsubsection{Envelope: state of matter}
\label{env_state}

\begin{figure}
\begin{center}
\begin{minipage}[c]{0.9\linewidth}
\includegraphics[width=\linewidth]{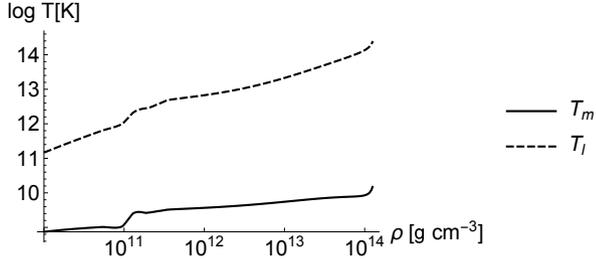}
\caption{\label{TlTm}
   For $T<T_m$ the ions crystallise, and for $T>T_l$ the ion plasma is
   well-approximated by a Boltzmann gas. In the intermediate
   temperature range, $T_m<T<T_l$, the ions form a Coulomb
   liquid. For the density range of interest to us, $\rho\gtrsim
   10^{10}\textrm{g cm}^{-3}$, we see that the proto-neutron star
   envelope is likely to be in a Coulomb-liquid state.}
\end{minipage}
\end{center}
\end{figure}

We know that the core-region entropy is
always dominated by the degenerate-neutron contribution, but the
envelope structure will change
depending on the temperature and density. In particular, for a given
density the envelope's ions are liable to form a Coulomb liquid at
lower temperatures, and an ideal Boltzmann
gas at higher temperatures, with the transition occurring at some
temperature $T_l$ . At the high temperatures we consider, shortly after birth, one
would not expect any part of the envelope to have cooled below the
temperature $T_m$ at which the ions freeze into a crystalline Coulomb
lattice, but we will also check this. Using the formulae from
Sect.\ref{sect:eos-compos-envel} we calculate the two transition
temperatures $T_l$ and $T_m$, plotting
the results in Fig. \ref{TlTm}.

We conclude that in the density and temperature range of interest to
us, the ions throughout the entire envelope are in a Coulomb-liquid
state. However, despite this, the Boltzmann-gas results for the thermal contributions are valid, as 
explained in Sect.\ref{sect:eos-compos-envel}.  We use these in the calculations which follow.

\subsubsection{Entropy contributions}

\begin{figure}
\begin{minipage}[c]{\linewidth}
\begin{center}
  \includegraphics[width=\linewidth]{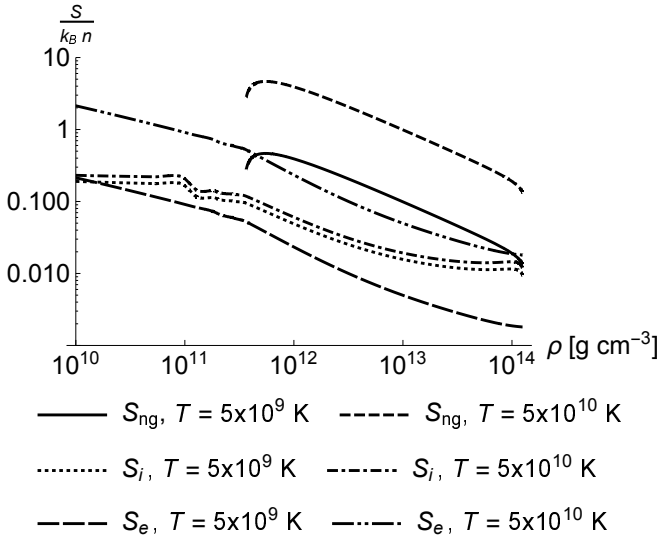}
\end{center}
\caption{\label{Sng_vs_ion}
  The entropy per baryon (in units of $k_B$) as a function of density,
  for the ion, electron and
  neutron-gas species, at (constant) temperatures of $5\times
  10^9,5\times 10^{10}$ K, as labelled.}
\end{minipage}
\end{figure}

Evaluating the relevant expressions from section
\ref{therm_EOS}, we plot in Fig. \ref{Sng_vs_ion} the entropy
  profiles for the neutron-gas, ion and electron components of the
  envelope. These are shown at two different constant temperatures, $T=5\times
  10^9$ and $5\times 10^{10}$ K. In the inner envelope we see that the
neutron-gas entropy is generally an order of magnitude larger than
the other two components, although at lower temperatures and higher
densities the ion entropy becomes comparable. In order to be able to approximate the
inner-envelope entropy by its neutron-gas component alone, we require
$T(\rho=\rho_{\rm cc})\gtrsim 10^{10}$ K. This is not a strong
restriction, though: cooler proto-NS
models are not likely to be of interest to us, since thermal
effects will become negligible; zero-temperature models will provide a
satisfactory description of the stellar structure.

\begin{figure}
\begin{center}
\begin{minipage}[c]{\linewidth}
\includegraphics[width=\linewidth]{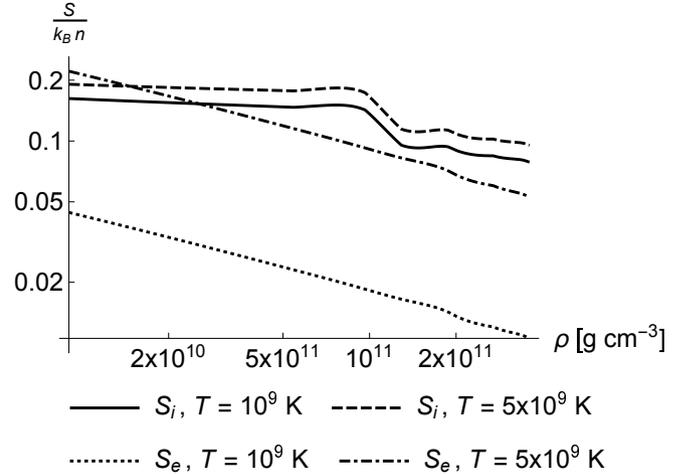}
\caption{\label{Se}
  Comparing the ion and electron entropy contributions (per baryon, in
  units of $k_B$) in the outer envelope. These are plotted as a function of
  density, at (constant) temperatures of $1$ and $5\times 10^9$ K, as labelled.}
\end{minipage}
\end{center}
\end{figure}

Below neutron-drip density there is no free
neutron gas, and its entropy therefore drops to zero at
$\rho=\rho_{\rm nd}$, leaving just the ion and electron
components. The latter clearly dominates at $T=5\times 10^{10}$ K, but
we do not expect the outer envelope to be this hot; see section
\ref{sect:isoT-vs-isoS}. We replot the ion and electron entropy
profiles in the outer envelope alone, and at constant temperatures of
$1$ and $5\times 10^9$ K, in Fig. \ref{Se}. For the latter temperature
the two components are comparable, whereas for the former the ion
entropy is a factor of $4-7$ bigger. Within our model we will be able
to choose the outer-envelope temperature, and should therefore ensure
it is relatively cool, so that the electron entropy can be neglected.

\begin{figure} 
\begin{center}
\begin{minipage}[c]{\linewidth}
\includegraphics[width=\linewidth]{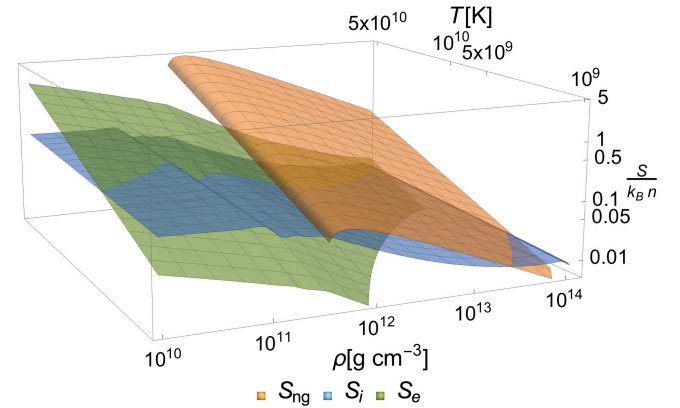}
\caption{\label{3D_ng_vs_ion}
  A three-dimensional version of figures \ref{Sng_vs_ion} and \ref{Se}, giving
  fuller information about the dependence of the ion, electron and neutron-gas
  entropy components as functions of density and temperature.}
\end{minipage}
\end{center}
\end{figure}

We conclude with a three-dimensional plot showing the behaviour
  of the three entropy components as a function of $T$ and $\rho$,
  giving a better qualitative picture of when different components
  dominate. We expect a realistic envelope temperature profile to decrease
  from $\gtrsim 10^{10}$ K at $\rho_{\rm cc}$ to $\lesssim 10^9$ K at
  $\rho=10^{10}\textrm{g cm}^{-3}$ (i.e. lines running from the far
  right to the near left of the plot). For such profiles, the electron
  entropy can be neglected throughout the envelope.

We have established that the star's entropy may reasonably be modelled
as due to a single dominant contribution in each region, as we had hoped: the
degenerate neutrons in the core, the neutron gas in the inner
envelope, and the ions in the outer envelope. This holds for the whole
temperature and density range of importance to us here.

\subsection{Our simplified thermal model}
\subsubsection{Isothermal vs isentropic}
\label{sect:isoT-vs-isoS}

In general the basic thermodynamic quantities have dependences
$S=S(T,\rho)$ and $T=T(S,\rho)$. Assuming that either $S$
or $T$ is constant in some region is very attractive, because it means
that the other one of the two quantities must become an explicit function of
$\rho$ alone, which makes it far easier to formulate the problem in a
manner suitable for an equilibrium solution.

From proto-NS simulations we see that the entropy per baryon $s_b$ in
the core becomes
approximately constant over very few seconds, whereas the temperature
varies by a large factor through this region \citep{BL86,pons99}. For this reason, we
will model the core as isentropic with some constant entropy per
baryon $s_{b0}$ (in units of $k_B$). Since the core makes up most of
the star, and will provide the dominant contribution to the star's
thermal pressure, we will treat this region first, and
choose a prescription for the envelope regions which matches to the
core. Therefore, the fundamental constant for defining the
thermal structure of a given stellar model will be $s_{b0}$.

The initial thermal evolution of the outer envelope, above the
neutrinospheres, is far faster than that of the neutrino-opaque
interior. Being transparent to neutrinos, this region cools rapidly
via $e^--e^+$ pair annihilation and plasmon decay and reaches an
isothermal state. In this work we will assume the outer-envelope temperature to take
some fixed value $T_{\rm oe}$ for all our models. Clearly we can
typically expect there to be a substantial jump between this value and the
temperature as calculated on the inner side of the envelope-core boundary
\beq
T_{\rm core}(\rho=\rho_{\rm cc})\gg T_{\rm oe}
\eeq
where matter continues to be heated by the trapped neutrinos.

We will therefore  need to construct a transition
region that leads us smoothly from the thermal structure of the outer
core to that of the outer envelope, similar to the approach employed
in \citet{Goussard97}. The simplest resolution to the 
problem -- given that we will need equations in closed form for our
iterative method (see section \ref{numerics}) -- is to
construct some simple closed-form function for either the entropy or
the temperature in the inner envelope, to match both to the core and
outer-envelope thermal structure. Experimenting with both possibilities, we have
found that prescribing $s_b$ in the inner envelope in terms of some
given function $s_{\rm ie}(\rho)$ and using this to
calculate $T$ leads to smaller errors than the other way around, and
so we adopt this approach.

In summary, then, our model for the thermal part of the equation of
state is the following:\\
(a) isentropic core, $\rho\geq \rho_{\rm cc}$, entropy
  due solely to degenerate neutrons;\\
(b) inner envelope, $\rho_{\rm nd}<\rho< \rho_{\rm cc}$, with entropy
 per baryon given by some fixed function $s_{\rm ie}(\rho)$, and $T$ calculated from
  this. Entropy is assumed to be due to the
  neutron-gas contribution alone;\\
  (c) isothermal outer envelope, $\rho\leq \rho_{\rm nd}$, with some fixed $T_{\rm oe}$, and entropy due
  to the ion contribution alone.\\

The exact functional forms of the temperature, entropy and thermal
pressure will be discussed in section \ref{Ptherm_sec}.
The model will not make sense once the typical internal temperature
drops to $T\sim T_{\rm oe}$, but by that point the thermal contribution to the
pressure, and hence to the magnetic-field distribution, will have
become negligible.

\subsection{Choosing outer-envelope temperature}

There appears to be very little discussion in the literature on the
temperature of the outer envelope.
\citet{Goussard97} took $T_{\rm oe}=0.2\ \textrm{MeV}=2.3\times 10^9$ K, without
providing any physical justification for this particular value. Studies on
proto-NS structure tend to use enclosed mass as the radial coordinate,
thus squashing the entire low-mass envelope into a very thin shell; no
detailed information can be gleaned from such plots. It is clear,
however, that the outer envelope cools earlier and faster than the
initially neutrino-opaque core -- and should therefore be assigned a
far lower temperature.

For simplicity, and for consistency with our neglect of the electron
entropy, we take $T_{\rm oe}=10^9$ K in all our models. 
Note that our results are almost 
completely independent of any choice less than roughly $10^{10}$ K; the outer envelope has little
influence on the structure of the star or its magnetic field. The
main rationale is to impose the expected substantial drop in
temperature between the core and
the outermost regions, and to avoid numerical issues related to finding a suitable
inner-envelope function to lead fairly smoothly between the core and
outer-envelope thermal structure (we take a quadratic in $\rho$, built assuming
the former is considerably bigger than the latter).

\section{Equilibrium equations}
\label{eqm_equations}

\subsection{Governing equations}  
  
Our model of the late stages of a proto-neutron star simply applies
the equations of magnetohydrodynamics to a rotating,
self-gravitating fluid body in equilibrium. The major novelty of our work is
the inclusion of a thermal-pressure term, which is conceptually simple
but complex in its details. To avoid additional difficulty we will work in Newtonian
gravity, even though a quantitative treatment of a NS should clearly
employ general relativity.

Firstly, the force balance in the star is
described by the Euler equation:
\beq \label{euler_orig}
-\frac{1}{\rho}\nabla P-\nabla\Phi+\nabla\Phi_r+\frac{1}{4\pi\rho}(\curl\bB)\times\bB=0,
\eeq
where $P$ is the (total) fluid pressure, $\Phi$ the
gravitational potential, $\bB$ the magnetic field and
\beq
\Phi_r=\frac{1}{2}r^2\sin^2\theta\Omega^2
\eeq
the rotational potential, with rigid rotation at frequency $\Omega$
assumed here.
The Euler equation is coupled to Poisson's equation:
\beq \label{poisson}
\nabla^2\Phi = 4\pi G\rho.
\eeq
We also need to satisfy the solenoidal constraint
\beq
\nabla\cdot\bB=0.
\eeq
The Euler equation has the same form as for previous studies of
zero-temperature neutron-star models. Here, however, the pressure has
two contributions: one from the degeneracy pressure (which is entirely
dominant in cold neutron stars), and a second thermal-pressure
term. We assume these two are separable, so that the total fluid
pressure is the sum of these, $P=P_0+P_{\rm th}$.

\subsection{Equilibrium equation of state}
\label{eqm_EOS}

The above system of equations is closed by an equation of state for
the stellar matter.  Models of matter in mature neutron stars generally posit an explicit
relation of the form
\beq
P=P(\rho)
\eeq
-- a barotropic equation of state, for which pressure is no longer an
independent variable. With the additional assumption of axisymmetry,
this leads to the magnetic field being described by a single PDE of one variable: the
Grad-Shafranov equation \citep{grad_rubin,shaf}.

On the other hand, \citet{reis09} argues that the
stratification of matter -- due to the presence of either thermal or
composition gradients -- means that the barotropic relation must be
abandoned; the pressure is no longer slave to the density, but can
have a more general dependence:
\beq\label{nonbaro}
P=P(\rho,x_\rmp,T,\dots).
\eeq
This removes a key step in deriving
the Grad-Shafranov equation, leading to additional terms that complicate the calculation of
equilibria. However, the result is typically wielded
in a far stronger way: to state that there is \emph{no restriction at
  all} on the magnetic field \citep{glam_lasky}, except the usual
solenoidal constraint. Were this to be true, the Grad-Shafranov equation could be
abandoned, and the magnetic field structure be chosen at will, with
the assumption that buoyancy would provide whatever force necessary to
satisfy the equilibrium condition -- as done by, e.g.,
\citet{mastrano} and \citet{akgun13}.
Whilst \citet{reis09} does argue for an upper limit, $B\sim
10^{17}$ G, to the ability of buoyancy forces to act in this way, none
of the models constructed in this manner make a quantitative treatment
of the effect of buoyancy forces or check, ex post facto, what kind of
force is being implicitly assumed to keep the star in equilibrium, and
whether it is consistent with physically-motivated equations of
state.

There is, however, another implicit assumption in equation
\eqref{nonbaro}, namely that reactions are slow enough that the
composition can be `frozen' in and act as an additional variable when
determining the pressure. 
If, however, reactions are fast enough, they will push the system towards beta equilibrium, and the proton
fraction (or however else the composition is quantified) will be a function
of $\rho$ alone, making the equation of state barotropic for this
purpose.

In the first stages of the proto-neutron star evolution, whilst the
matter is still opaque to neutrinos, the beta equilibration timescale
$\tau_\beta\ll \tau_A$ except in a very thin shell close to the
surface, where the two timescales are comparable \citep{camelio17}, and the star can always be considered to be in beta
equilibrium for our analysis. Once the core has cooled below $T\approx
5\times 10^{10}$ K \citep{pons99} and has become
transparent to neutrinos, we can assume that standard modified Urca
reactions act to restore beta equilibrium, leading to a
timescale \citep{villain05}
\beq
\tau_\beta \approx 0.5 \left(\frac{T}{10^{10}\; \mbox{K}}\right)^{-6}
                               \left(\frac{\rho}{\rho_0}\right)^{1/3}\left(\frac{x_\rmp}{0.01}\right)^{1/3}\; \mbox{s}\, ,\label{tau}
\eeq
where $\rho_0$ is the nuclear saturation density. If nucleonic or
hyperonic direct Urca reactions are possible the equilibration
timescale will be even shorter, so in general for temperatures
$T>10^{10}$ K, such as those we consider in our model, reactions will
occur on a faster timescale than the magnetic field can
adjust. It is therefore safe to assume that an unmagnetised proto neutron star is
  in chemical equilibrium (although not, at this early stage, in thermal equilibrium).

The presence of the magnetic field introduces a considerable
  complication to this discussion. To see this, it is enough to consider a cold NS
model composed of neutron, proton and electron fluids and satisfying
local charge neutrality $n_{\rm p}=n_{\rm e}$, for which one can show
(e.g. \citet{LAG}) that:
\beq\label{chem_eqm_B}
\nabla\brac{\mu_{\rm p}+\mu_{\rm e}-\mu_{\rm n}}
=\frac{1}{n_{\rm p}}(\curl\bB)\times\bB.
\eeq
When the right-hand side is zero, this reduces to a statement that the
fluids are in chemical equilibrium. This equation, therefore, seems to
imply that a magnetic field will generally drive a star out of
chemical equilibrium by a (small) amount scaling with
$B^2$. Furthermore, using a toy model
  of a thin fluxtube rising through hot unmagnetised neutron-star matter,
  \citet{reis09} argued that beta re-equilibration also becomes considerably
  slower in the presence of a magnetic field. \citet{gus17} later
    reached the same conclusion through analysis
    of the evolution equations for a magnetised multifluid
    star. They argue that approximations valid for an unmagnetised
    star are no longer appropriate, and find equilibration
    timescales of the order of $10^6$ yr for the kinds of temperature
    and field strength we consider here -- clearly vastly longer than
    the duration of the late proto neutron star phase.

If this view is correct, a magnetic field initially out of chemical
  equilibrium will remain so throughout this early phase, experiencing
  an additional buoyancy force that is absent from our modelling. The
  further the magnetised star is from equilibrium, the less reliable our models will be; in the
  extreme, the (presumably) complex smallscale field generated at
  birth could be preserved into the mature phase of the star,
  although we find it more likely that the magnetic field would
    never take the star far out of chemical equilibrium. Whatever the
  result of the birth physics, however, in no case
  would it be appropriate to model this phase simply by `choosing' an arbitrary
  magnetic-field configuration and invoking buoyancy forces to balance it.

In principle the interplay between a
magnetic field and chemical reactions could be studied by some
future nonlinear numerical evolutions of the hot, magnetised multifluid
neutron star, hopefully providing a definitive resolution of the
issue. In the absence of such a study, we regard our approach as a
sensible start. If nothing else, our results represent whatever
  subclass of proto-neutron-star magnetic fields that do not drive the
  star out of beta equilibrium. If a magnetised neutron star leaves the birth
  phase already close to chemical equilibrium, such models may be
  accurate enough.

Returning to the pressure-density relation and with our assumption that the pressure is separable, we may very
generally write:
\beq
P(\rho,x_\rmp,s,T)=P_0(\rho,x_\rmp)+P_{\rm th}(\rho,s,T,x_\rmp).
\eeq
However, in chemical equilibrium $x_\rmp=x_\rmp(\rho)$, and since
any convective circulation of
matter will already have ceased, we can expect
thermodynamic quantities to be constant along isopycnic contours,
i.e. $T(\rho),s_b(\rho)$. In this work we will adopt a model where
either $s_b$ or $T$ is a prescribed function of $\rho$ in each region, so that the
other quantity may then be calculated from this, and will also be a
function of $\rho$. As a result, finally,
we find that the equation of state will return to being barotropic:
\beq
P=P(\rho,T(s(\rho)),s(T(\rho)))=P(\rho).
\eeq

As the star cools below $T\approx 10^{10}$ K, the reaction and Alfven
crossing timescales become comparable, and deviations from chemical
equilibrium and magnetic effects can balance each other, as predicted
by equation \eqref{nonbaro}. However the strong dependence on
temperature of the timescale in (\ref{tau}) means that the star
rapidly enters the frozen regime, in which the field simply adapts to
the fluid configuration as it relaxes. We thus expect any modest
deviation from a barotropic equilibrium to be washed out on an
equilibration timescale of a few minutes as the star cools, unless
some other physical mechanism is at work to maintain this
out-of-equilibrium state (see, e.g., \citet{ofen_gus} and
\citet{castillo} for some recent modelling of magnetic-field evolution
during this later phase). Note that our conclusions only apply to
quasi-stationary situations: thermal or composition gradients are clearly
important actors in proto-NS dynamics, such as the study of
oscillation modes.

For the cold part of the EOS, the majority of studies of NS magnetic equilibria
have assumed a single polytrope to govern the pressure-density
relation, which is a poor reflection of the real star (see however
\citet{KKKK} for an exception to this). As a minimal, but physically
well-motivated, extension to this, we will take a two-piece polytropic equation of
state, with the core and envelope regions having different polytropic
indices:
\beq\label{piece_poly}
P=\begin{cases}
        k_1\rho^{1+1/N_1}\textrm{ in envelope, i.e. }\rho<\rho_{\rm cc}\\
        k_2\rho^{1+1/N_2}\textrm{ in core, i.e. }\rho\geq\rho_{\rm cc}.
      \end{cases}
\eeq
Continuity and thermodynamic consistency mean that $k_1$ and $k_2$ are
not independent of one another \citep{read09}.

\section{The thermal pressure}
\label{Ptherm_sec}

\subsection{Non-dimensionalising}

For zero-temperature stellar models in Newtonian gravity, it is
natural to use combinations of $G$, the central density $\rho_c$ and the
equatorial radius $R_*$ to make all quantities dimensionless. This
both simplifies the solution method and ensures that quantities are
all (very broadly) of order unity, which decreases the error in
numerical calculations. Because
quantities like the polytropic constant $k$ drop out from the
dimensionless solution, 
results can be rescaled to any desired stellar model by using the
requisite values of $\rho_c$ and $R_*$ to restore the dimensions $\sfM,\sfL,\sfT$.

Now that we have thermal quantities, however, we need an extra quantity including
the temperature dimension $\mathsf{O}\!\!\!\! -$, in order to make everything
dimensionless. We will find that results for hot models will no longer
be rescalable.

Note that for supernovae and proto-neutron stars it is convenient to
work with the entropy per baryon $s_b$, in units of the Boltzmann
constant $k_B$. $s_b$ is dimensionless and is related to the entropy
density $S$ through:
\beq
s_b=\frac{\mathcal{S}}{k_B\mathcal{N}}=\frac{\mathcal{S}/V}{k_B\mathcal{N}/V}=\frac{S}{k_B n},
\eeq
where $\mathcal{S}$ is the entropy and $\mathcal{N}$ the number of
particles within some region.
Let us use $k_B/m_{\rm u}$ as our fourth quantity for
nondimensionalising; we see it includes the desired dimension of
temperature, since
\beq
[k_B/m_{\rm u}]=[\mathcal{S}]\sfM^{-1}=\sfL^2\sfT^{-2}\mathsf{O \!\!\!\! -}^{-1}.
\eeq
The dimensionless entropy density is then given by:
\beq
\hat{S}=\frac{S}{(k_B/m_{\rm u})\rho_c}.
\eeq
Conveniently, $S/\rho$ in dimensionless units is then given by:
\beq
\frac{\hat{S}}{\hat\rho}
=\frac{S m_{\rm u}}{k_B\rho_c}\frac{\rho_c}{\rho}=\frac{S m_{\rm u}}{k_B\rho}=s_b,
\eeq
the entropy per baryon in units of $k_B$.
Next we need to find the temperature in dimensionless units. Using the
dimensions of $k_B/m_{\rm u}$ above, and the fact that
\beq
[G\rho_c]=\sfT^{-2},
\eeq
we see that the following combination of quantities has dimensions of
temperature:
\beq
(k_B/m_{\rm u})^{-1}G\rho_c R_*^2
\eeq
and so
\beq\label{dimless_T}
\hat{T}=\frac{k_B T}{m_{\rm u} G\rho_c R_*^2}.
\eeq
We are using $m_{\rm u}$ (the atomic mass unit) instead of the neutron mass,
although our core model only considers the thermal contribution of the
neutrons.

Not everything in dimensionless form can be rescaled at will,
however. The Fermi temperature $T_{\textrm F}$ depends on the Fermi energy $\varepsilon_{\textrm F}$,
which in turn depends on the physical value of density within the star
(not just at the centre). This
means that we will not be able to remove all physical quantities from
our unit system (even though we \emph{have} got rid of $G$ and
$k_B/m_{\rm u}$). We will see that this is not a problem for obtaining
solutions, but it does mean we will 
have to specify some stellar quantities in advance, in physical units. In
fact, even cold models with our new EOS will be specific to one
stellar model, since at least two densities enter the calculation: at the centre
and at the transition between different adiabatic
indices. We are still able to choose dimensionless units such that
$\hat\rho=\rho/\rho_c$, but having any kind of internal transition at
some given \emph{physical} density clearly means the physical central density
must be specified in our dimensionless scheme.

\subsubsection{Core}

In order to find the thermal-force scalar $\Theta$ we first need an
expression for the thermal pressure in convenient, dimensionless
form. Comparing the expressions for $P_{\rm th}$ and $S$ from section \ref{sect:therm.core},
we see that
\beq\label{Pth_core}
P_{\rm th}=\frac{1}{3}ST.
\eeq
In dimensionless units,
\beq
\hat{P}_{\rm th}=\frac{1}{3}\frac{\hat{S}}{\hat\rho}\hat\rho\hat{T}
 =\frac{1}{3} s_b\hat\rho\hat{T}.
 \eeq
To use this expression, we need to know the relation between $s_b$ and
$\hat{T}$. From \ref{sect:therm.core} we see that
\beq\label{s-T_core}
s_b=\frac{\hat{S}}{\hat\rho}=\frac{\pi^2}{2}\frac{k_B T}{\varepsilon_{\textrm F}},
 \eeq
 using $n=\rho/m_{\rm u}$ and $\varepsilon_{\textrm F}=k_B T_{\textrm F}$. Now, the Fermi energy is given by
\begin{align}
\varepsilon_{\textrm F}
&=58.44\brac{\frac{n}{n_0}}^{2/3}\textrm{ MeV}\nn\\
&=9.363\times 10^{-5}\brac{\frac{\rho}{\rho_c}}^{2/3}\brac{\frac{\rho_c}{\rho_{\rm nuc}}}^{2/3}\textrm{ erg}\nn\\
&=9.363\times 10^{-5}\hat\rho_{\rm nuc}^{-2/3}\hat\rho^{2/3}\textrm{ erg},
\label{epsF_core}
\end{align}
where $\rho_{\rm nuc}=2.8\times 10^{14}\textrm{g cm}^{-3}$ is nuclear mass density.
Using the above expression for $\varepsilon_{\textrm F}$ in equation
\eqref{s-T_core} and rearranging, we see that the
dimensionless temperature is given by
\begin{align}
\hat{T} &= \frac{k_B}{m_{\rm u} G\rho_c R_*^2}\frac{2}{\pi^2}\frac{\varepsilon_{\textrm F}}{k_B}s_b\nn\\
 &= 0.1712\brac{\frac{\rho_c}{10^{15}\textrm{ g cm}^{-3}}}^{-1}\brac{\frac{R_*}{10^6\textrm{ cm}}}^{-2}
               \hat\rho_{\rm nuc}^{-2/3} s_b\hat\rho^{2/3}.
\label{core_T}
\end{align}
It was not necessary to specify $\rho_c$ in advance for earlier,
zero-temperature equilibria \citep{tomi_eri,LJ09}, but since we
need the ratio $\rho_{\rm nuc}/\rho_c$ here, it is clear that we must now
work with the central density in physical units for each model.

We now turn to the thermal-pressure force. This takes the dimensionless form:
\beq
\nabla\hat\Theta=\frac{\nabla\hat{P}_{\rm th}}{\hat\rho}=\frac{1}{3\hat\rho}\nabla(\hat\rho\hat{T}s_b).
\eeq
Since we know that $\hat\Theta=\hat\Theta(\hat\rho)$, the thermal force
above may be written, using the chain rule, as
\beq
\nabla\hat\Theta(\rho) =\hat\Theta'\nabla\hat\rho
 = \frac{1}{3\hat\rho}\brac{\hat\rho\hat{T}s'_b+\hat\rho\hat{T}'s_b+\hat{T}s_b}\nabla\hat\rho
\eeq
where all primes denote differentiation with respect to
$\hat\rho$. Next, note that $\nabla\rho\neq 0$ throughout the star
except for its exact centre, and that at the centre itself
we must have $\nabla\Theta=0$ for regularity. Therefore we may cancel
the $\nabla\hat\rho$ terms from the LHS and RHS of the above
expression and integrate, to yield
\beq\label{full_Theta}
\hat\Theta=\frac{1}{3}\int\brac{\hat{T}s'_b+\hat{T}'s_b+\frac{\hat{T}s_b}{\hat\rho}}\ \rmd\hat\rho,
\eeq
which will include some integration constant. There should be freedom
in choosing this constant, since the physical equilibrium of the star
depends only on $\Theta'(\rho)$ and not $\Theta$ itself; it is like a
gauge freedom. This will be useful to us later.

To evaluate $\Theta$ in general, we would need both $S$ and $T$ inputs
from numerical simulations of the late proto-neutron-star
phase. However, as discussed in \ref{sect:isoT-vs-isoS} , we will adopt a
simplified model that avoids this requirement. In our model the core
is assumed to be isentropic:
\begin{align}
s_b &=s_{b0}=\frac{\hat{S}}{\hat\rho}\\
\hat{T} &= 0.1697\brac{\frac{\rho_c}{10^{15}\textrm{ g cm}^{-3}}}^{-1}\brac{\frac{R_*}{10^6\textrm{ cm}}}^{-2}
               \hat\rho_{\rm nuc}^{-2/3} s_{b0}\hat\rho^{2/3}.
\label{T_isentrop}
\end{align}
We now have $T$ as a function of $\rho$, and so
\beq
\hat\Theta=\frac{1}{3}\int\brac{\hat{T}'s_b+\frac{\hat{T}s_b}{\hat\rho}}\ \rmd\hat\rho,
\eeq
using the isentropic assumption $s'_b=0$.
Integrating the above, we have
\beq\label{Theta_core}
\hat\Theta=\frac{5}{6} s_{b0}\hat{T}.
\eeq
We now move on to calculating the thermal-pressure force in the
envelope regions for our model.

\subsubsection{Outer envelope}

As discussed in section \ref{sect:isoT-vs-isoS},  we take the outer envelope to be
isothermal with some fixed physical temperature $T_{\rm oe}$ in kelvin for all stellar
models (except the zero-temperature models we compare with in the
results section). We first convert this to a non-dimensional value
using equation \eqref{dimless_T}; unlike the physical value, this will vary between
models.

We have established that the thermal structure is dominantly due to
the ions. The dimensionless thermal pressure for a
Boltzmann gas of ions is
\beq
\hat{P}_{\rm th}=\frac{n_{\rm i} k_B T}{G\rho_c^2 R_*^2}=\frac{\hat{\rho}\hat{T}}{A},
\eeq
using the fact that $A'=A$ in the outer envelope. From the above we
calculate the form of the thermal-pressure scalar:
\begin{align}
\hat\Theta &= \int\frac{1}{\hat\rho}\td{\hat{P}_{\rm th}}{\hat\rho}\ \rmd\hat\rho\nn\\
  &= \frac{1}{A}\int \brac{\frac{\hat{T}}{\hat\rho}+\td{\hat{T}}{\hat\rho}}\ \rmd\hat\rho,
\end{align}
where we have assumed for simplicity that $A$ is constant in the
region of interest to us (i.e. the inner part of the outer envelope);
we take $A=100$ as a representative value for this region.

For an isothermal envelope we then have
\beq\label{Theta_oe}
\hat\Theta=\frac{\hat{T}_{\rm oe}\ln\hat\rho}{A}.
\eeq
This is only evaluated within the star and not at the surface, so no
issues arise from the divergent nature of this expression for $\rho\to
0$.

For the purposes of the equilibrium calculation, the ion entropy is not
used explicitly to derive the thermal-pressure scalar. We will however
need it in order to construct an entropy function for the inner
envelope; see next. Equation \ref{eq:mu.S.ions} gives the
entropy density per ion. To convert to an entropy per baryon (in units
of $k_B$), we divide the dimensionless form of this entropy by
$\hat\rho$, to give: 
\begin{align}
  s_b=\frac{\hat{S}_{\rm i}}{\hat\rho}
  &= \frac{1}{A}\left[\frac{5}{2}-\ln(n_{\rm i}\lambda_{\rm i}^3)\right]\nn\\
  &= \frac{1}{100}\left[\frac{5}{2}
         -\ln\brac{ 4.47\times 10^{-5}\rho_{c,15}^{-1/2} R_6^{-3}}
         -\ln\brac{\hat\rho \hat{T}^{-3/2}}\right],
\end{align}
where we evaluate the expression for $A=100$ on the second line.

\subsubsection{Inner envelope}

With the above expression, we evaluate the entropy per baryon at the inner edge of the outer
envelope $s_{\rm oe}$. The value in the core is also known, and
fixed at $s_{b0}$ from the outset of the calculation. We now construct
a quadratic function $s_{\rm ie}$ for the entropy per baryon in the
inner envelope to lead between these two values:
\beq
s_{\rm ie}(\hat\rho)
 =s_{b0}-\frac{(s_{b0}-s_{\rm oe})}{(\hat\rho_{\rm cc}-\hat\rho_{\rm nd})^2}
                                                        (\hat\rho_{\rm cc}-\hat\rho)^2.
\eeq
Using this prescription gives us models where $s_b$ throughout the star is
continuous and quite smooth.

In the inner envelope we assume that only the neutron gas contributes
to the thermal structure. This is physically the same as in the core,
except that number density factors $n$ are weighted with a prefactor
\beq
\xi\equiv\frac{X_{\rm n}}{(1-u)}
\eeq
that determines the microscopic density of the degenerate neutron gas outside nuclei  within the
region.
The thermal pressure takes the same form as in the core, equation
\eqref{Pth_core}, except that the relation between the entropy and the
temperature is now given by
\beq
s_b=s_{\rm ie}(\hat\rho)=\frac{\hat{s}_{\rm ng}}{\hat\rho}=\frac{\pi^2}{2}\xi\frac{k_B T}{\varepsilon_{\textrm F}}.
\eeq
Through its density term, the Fermi energy \eqref{epsF_core} in the inner envelope picks up a prefactor $\xi^{2/3}$.
Rearranging for the dimensionless temperature as in the core case, then, we
arrive at the same expression as equation \eqref{core_T}, but with
a prefactor of $\xi^{-1/3}$:
\beq
\hat{T} = 0.1712\brac{\frac{\rho_c}{10^{15}\textrm{ g cm}^{-3}}}^{-1}\brac{\frac{R_*}{10^6\textrm{ cm}}}^{-2}
               \hat\rho_{\rm nuc}^{-2/3} \xi^{-1/3}s_b\hat\rho^{2/3}.
\eeq
The presence of $\xi$ is a considerable complicating factor in our
calculation. In problems where a quantitative treatment of inner-envelope physics 
is required (e.g. for pulsar glitches), it is clearly important. Note,
however, that for almost the whole density range of the inner envelope
$\xi\approx 0.6-0.8$, corresponding to a prefactor $\xi^{-1/3}\approx
1.1-1.2$ in the above equation. Only in the region $3.5\times
10^{11}<\rho[\textrm{g cm}^{-2}]\lesssim 2\times 10^{12}$ does $\xi$
have larger variation -- but this corresponds to, at most, a very few
grid points for us.

We have experimented with different prescriptions for $\xi$, finding
that the mismatch at the envelope-core boundary for $\xi\neq 1$
introduces considerable error (in the sense of not satisfying the
virial test of section \ref{virial} to high precision), but with imperceptible changes in the
actual stellar models. For this reason we will make the
simplification, from now on,
that $\xi=1$ throughout the inner envelope. Since the entropy is a
prescribed function $s_{\rm ie}(\hat\rho)$, the inner-envelope
temperature can then be calculated from
\begin{align}
\hat{T} &= 0.1712\brac{\frac{\rho_c}{10^{15}\textrm{ g cm}^{-3}}}^{-1}\brac{\frac{R_*}{10^6\textrm{ cm}}}^{-2}
               \hat\rho_{\rm nuc}^{-2/3} s_{\rm ie}(\hat\rho)\hat\rho^{2/3}\nn\\
 &\equiv \mathfrak{c} s_{\rm ie}(\hat\rho)\hat\rho^{2/3},
\end{align}
where we have defined the constant $\mathfrak{c}$ to absorb the
various numerical prefactors above.
We may now calculate the thermal-force scalar in the inner envelope:
\begin{align}
\hat\Theta &= \frac{1}{3}\int\brac{\hat{T}s'_{\rm ie}+\hat{T}'s_{\rm ie}+\frac{\hat{T}s_{\rm ie}}{\hat\rho}}\ \rmd\hat\rho\nn\\
           &= \frac{\mathfrak{c}}{3}
                    \left[ s_{\rm ie}^2\hat\rho^{2/3} + \int s^2_{\rm ie}\hat\rho^{-1/3}\ \rmd\hat\rho  \right]\nn\\
           &= \frac{\mathfrak{c}\hat\rho^{2/3} s_{\rm ie}}{3}
                   \left[s_{\rm ie}+\frac{f(\hat\rho)}{s_{\rm ie}}\right]\nn\\
           &= \frac{\hat{T}}{3}\left[s_{\rm ie}+\frac{f(\hat\rho)}{s_{\rm ie}}\right],
\label{Theta_ie}
\end{align}
where $f(\hat\rho)$ is a rather messy quartic in $\hat\rho$ emerging
from the above integration.

\subsubsection{Matching $\Theta$ contributions}

The freedom to choose the integration constant for $\Theta$ in each
region means we are able to adjust these to produce a continuous,
quite smooth, $\Theta$ profile from the centre to the surface of the
star. In particular, let $\Theta_{\rm core}, \Theta_{\rm ie},
\Theta_{\rm oe}$ denote the functions from equations
\eqref{Theta_core}, \eqref{Theta_ie}, \eqref{Theta_oe} without integration constants added on. For
the core we choose $\Theta=\Theta_{\rm core}$, i.e. without
integration constant. We then move to the inner envelope, creating a
function $\tilde\Theta_{\rm ie}$ that matches to $\Theta_{\rm core}$
at the envelope-core boundary, and then create an adjusted outer-envelope
function $\tilde\Theta_{\rm oe}$ to match to this $\tilde\Theta_{\rm ie}$ at the inner-outer
envelope boundary. For all points with $\rho=0$, $\Theta$ is taken to
have the value $\tilde\Theta_{\rm oe}^{\rm surf}$ obtained from
evaluating $\tilde\Theta_{\rm oe}$ at the last gridpoint within the
star. To summarise, then,
\beq
\Theta(\rho)=\begin{cases}
  \Theta_{\rm core}(\rho) & \rho\geq\rho_{\rm cc}\\
 \Theta_{\rm ie}(\rho)-\Theta^{\rm cc}_{\rm ie}+\Theta^{\rm cc}_{\rm core}
\equiv \tilde\Theta_{\rm ie}(\rho)
 & \rho_{\rm nd}<\rho<\rho_{\rm cc}\\
  \Theta_{\rm oe}(\rho)-\Theta^{\rm nd}_{\rm oe}+\tilde\Theta^{\rm nd}_{\rm ie}
\equiv \tilde\Theta_{\rm oe}(\rho)
  & 0<\rho\leq\rho_{\rm nd}\\
  \tilde\Theta_{\rm oe}^{\rm surf}
  & \rho=0
\end{cases}
\eeq
where the superscripts cc and nd denote quantities evaluated at
$\rho=\rho_{\rm cc}$ and $\rho_{\rm nd}$ respectively.

\section{Numerical solution method}
\label{numerics}

A large number of previous studies have solved for magnetic-field
equilibrium models in NSs using numerical iterative schemes; of
particular note are the first solutions for a linked poloidal-toroidal
field in Newtonian gravity \citep{tomi_eri} and full general
relativity \citep{uryu19}. As in \citet{tomi_eri}, we will employ the
Hachisu self-consistent field (HSCF) method \citep{hachisu}, a robust
iterative procedure that has 
several advantages over perturbative methods: one can solve for models
up to Keplerian rotation rates, with extremely strong magnetic
fields, and include the contributions from high multipoles in the
solution without significant extra difficulty. The resulting models are true
self-consistent equilibria; whilst perturbative studies account for
the effect of the fluid distribution on the magnetic field, a
numerical iterative method is also able to account for the
back-reaction of the field on the fluid.
Complementary to these magnetised models, there has been a limited
amount of research on the construction and use of self-consistent
methods to build hot, rotating and unmagnetised stellar models
\citep{jackson05,Goussard97,camelio19}.
We will build on this body of work to produce models of hot NSs with
magnetic fields.

The HSCF method is semi-analytic, in that it exploits certain
closed-form expressions for the fluid and magnetic field in order to
iterate towards an equilibrium solution. These are valid for cold
polytropic stellar models; in the following we check whether they can be
adapted for models with a more realistic
description of the pressure in a hot proto neutron star: including
both a model of the thermal pressure, and a
piecewise-polytropic description of the 
degeneracy-pressure profile.

\subsection{Iterative solution: the fluid distribution}

\subsubsection{First-integral form of the Euler equation}

Firstly, we need to be able to write the Euler equation in integral
form. We have argued that the zero-$T$ part of the EOS is barotropic,
$P_0=P_0(\rho)$, meaning that one can write:
\beq
\frac{\nabla P_0}{\rho}=\nabla H,
\eeq
where $H$ is the enthalpy per unit mass, and is found from the integral:
\beq\label{H_int}
H = \int\limits_0^{P_0}\frac{\rmd \tilde{P}}{\rho(\tilde{P})},
\eeq
where the tildes denote dummy integration variables.
Here we have used the enthalpy, whereas some other papers use the
chemical potential per unit mass,
\beq
\tilde\mu\equiv\frac{\mu}{m}.
\eeq
Provided that we are able to separate out
thermodynamic quantities into zero- and finite-temperature pieces, the
two are equivalent. We can see this from the Euler relation:
\begin{align}
\tilde\mu=\frac{\mu}{m}=\frac{U+P-sT}{mn}
&= \frac{1}{\rho}(U_0+P_0+U_{\rm th}+P_{\rm th}-sT)\nn\\
&= H_0 + \frac{1}{\rho}(U_{\rm th}+P_{\rm th}-sT),
\end{align}
using the definition of $H$. Thus, at zero temperature there is no
distinction between $H$ and $\tilde\mu$.

Using $H$, the Euler equation for a cold star becomes:
\beq
\nabla(H+\Phi-\Phi_r)=\frac{1}{4\pi\rho}(\curl\bB)\times\bB.
\eeq
Finally, if we take the curl of this we see that
\beq
\curl\left[\frac{1}{4\pi\rho}(\curl\bB)\times\bB\right]=0\implies \frac{1}{4\pi\rho}(\curl\bB)\times\bB=\nabla M
\eeq
for some scalar function $M$. We then arrive at an important result
for the HSCF scheme: that the Euler equation becomes a Bernoulli
equation, and therefore may immediately be expressed in
first-integral form:
\beq\label{int_Euler_cold}
H+\Phi-\Phi_r-M=C,
\eeq
where $C$ is an integration constant, which is fixed through boundary
conditions at the surface.

The thermal quantities $T$ and $S$
are also functions of $\rho$ in our model, and as a result the thermal
pressure force may be written
\beq
\frac{\nabla P_{\rm th}}{\rho}
=\nabla\Theta.
\eeq
As a result, the Euler equation \eqref{int_Euler_cold} may trivially
be generalised to:
\beq\label{int_Euler_hot}
H+\Theta+\Phi-\Phi_r-M=C.
\eeq
Now, using the explicit form
of $\Phi_r$, we have:
\beq
H+\Theta+\Phi-\frac{\Omega^2}{2}r^2\sin^2\theta-M=C,
\eeq
and we \emph{define} the surface as being where
\beq
H=0.
\eeq
It would be most natural to define it as being where the density
drops to zero instead, but this is not convenient for numerical
implementation. So, evaluating the above Euler equation at the polar
and equatorial surfaces -- in code units where the equatorial surface
is at $\hat{r}_{\rm eq}=\hat{R}_*=1$ -- we have
\begin{align}
\hat\Theta(\hat{r}_{\rm pole})+\hat\Phi(\hat{r}_{\rm pole})-\hat{M}(\hat{r}_{\rm pole}) &= \hat{C},\\
\hat\Theta(1)+\hat\Phi(1)-\frac{\hat\Omega^2}{2}-\hat{M}(1) &= \hat{C},
\label{intC}
\end{align}
Subtracting the second equation from the first gives us an expression
for the rotation rate:
\beq\label{Om2}
\hat{\Omega}^2=2[\hat{\Phi}(1)-\hat{\Phi}(\hat{r}_{\rm pole})]
                             -2[\hat{M}(1)-\hat{M}(\hat{r}_{\rm pole})],
\eeq
where the $\hat\Theta$ terms cancel, since the  function is constant
along any density contour (in this case, the $\hat\rho=0$ contour).
Now that we have $\hat\Omega$ we may also use either of the above
boundary equations to calculate $C$.

\subsubsection{Iterative method: second key step}

The second important requirement of the HSCF method is the ability to
find a closed-form inversion for $\rho=\rho(H)$. This is only true for
particular special choices of the EOS, like a polytrope:
\beq
P=k\rho^\gamma=k\rho^{1+1/N}.
\eeq
For this polytropic EOS a straightforward integration, using equation
\eqref{H_int}, shows that
\beq \label{H_of_rho}
H=k(1+N)\rho^{1/N},
\eeq
which can be rearranged to give
\beq \label{rho_of_H}
\rho=\brac{\frac{H}{k(1+N)}}^N.
\eeq
This is effectively the iterative step for the method, used to find a
new density distribution -- one closer to an equilibrium state than the
previous one.

Now, if we work in dimensionless units by dividing all
physical quantities by combinations of the central density $\rho_c$,
stellar radius $R_*$ and the gravitational constant $G$, we can make
the expressions even simpler. Evaluating \eqref{H_of_rho} at the centre of the star in
dimensionless units, we
have:
\beq
\hat{H}_c=\hat{k}(1+N)\hat\rho_c^{1/N}=\hat{k}(1+N),
\eeq
where hats denote dimensionless variables, and where we have used
$\hat\rho_c=1$. Now substituting this relation into equation
\eqref{rho_of_H}, we get the very simple result:
\beq\label{iterate_poly}
\hat\rho=\brac{\frac{\hat{H}}{\hat{H}_c}}^N.
\eeq
We have eliminated the polytropic constant $k$ by working in
dimensionless units. This means that the final dimensionless model may be
redimensionalised to a whole set of models with different $k$, meaning
different mass and radius.

\subsubsection{Piecewise polytrope}

We now generalise the above result to the two-piece polytrope of equation
\eqref{piece_poly}. The relevant basic formulae for a relativistic
multi-piece polytrope are given in \citet{read09}. Because we work in Newtonian
gravity, however, we have amended the expressions of \citet{read09} to
remove the relativistic term $\rho c^2$ from the energy density.

Firstly, the requirement that the pressure should be continuous across the
boundary between the two polytropes means that the two polytropic
constants may not be chosen independently. In particular, the
internal energy $U(\rho)$ and enthalpy $H(\rho)$ are given by:
\beq
U(\rho)
 =\left[ \frac{U(\rho_{i-1})}{\rho_{i-1}}-\frac{k_i}{(\gamma_i-1)}\rho_{i-1}^{\gamma_i-1} \right]\rho
    + \frac{k_i}{(\gamma_i-1)}\rho^{\gamma_i},
\eeq
    
\beq
H(\rho)
 =\frac{U(\rho_{i-1})}{\rho_{i-1}}-\frac{k_i}{(\gamma_i-1)}\rho_{i-1}^{\gamma_i-1}
     + \frac{k_i\gamma_i}{(\gamma_i-1)}\rho^{\gamma_i-1}.
\eeq

Since we consider a two-piece polytrope, the transition
densities are $\rho_{01}=0$ (the stellar surface) and $\rho_{12}=\rho_{\rm cc}$
(the envelope-core transition). Continuity of pressure at the
envelope-core boundary:
\beq
P=k_1\rho_{\rm cc}^{1+1/N_1}=k_2\rho_{\rm cc}^{1+1/N_2}
\eeq
immediately gives
\beq
k_1=k_2\rho_{\rm cc}^{1/N_2-1/N_1}.
\eeq
Since $U/\rho \to 0$ at the surface, we find for the envelope:
\beq
U_{\rm env}(\rho)
= \frac{k_1}{(\gamma_1-1)}\rho^{\gamma_1}
= k_2\rho_{\rm cc}^{1/N_2-1/N_1} N_1\rho^{1+1/N_1},
\eeq

\begin{align}
H_{\rm env}(\rho)
= \frac{k_1\gamma_1}{(\gamma_1-1)} \rho^{\gamma_1-1}
&= (N_1+1) k_1\rho^{1/N_1}\nn\\
&= (N_1+1) k_2\rho_{\rm cc}^{1/N_2-1/N_1}\rho^{1/N_1}.
\end{align}

The $U_{\rm env}$ will not be directly used in our solution,
but is needed in
deriving the enthalpy for the core region:
\beq
H_{\rm core}(\rho)
= k_2\left[ (N_1-N_2)\rho_{\rm cc}^{1/N_2}+(N_2+1)\rho^{1/N_2} \right].
\eeq
The central enthalpy, in code units, is therefore
\beq\label{Hc}
\hat{H}_c(\rho)
= \hat{k}_2\left[ (N_1-N_2)\hat\rho_{\rm cc}^{1/N_2}+N_2+1\right].
\eeq

Now dividing the enthalpy by its central value allows us to eliminate
explicit mention of the polytropic constant, as in the
single-polytrope case. This leads to an inversion of $\hat\rho$ in
terms of $\hat{H}$ with different forms in each region, as follows:
\beq\label{iter_rho_env}
\hat\rho_{\rm env}=\hat\rho_{\rm cc}
\left\{
  \frac{\left[ N_1-N_2+(N_2+1)\hat\rho_{\rm cc}^{-1/N_2} \right]}{(N_1+1)}
  \frac{\hat{H}_{\rm env}}{\hat{H}_c} 
\right\}^{N_1},
\eeq

\beq\label{iter_rho_core}
\hat\rho_{\rm core}=\hat\rho_{\rm cc}
\left\{
  \frac{\left[ N_1-N_2+(N_2+1)\hat\rho_{\rm cc}^{-1/N_2} \right]}{(N_2+1)}
  \frac{\hat{H}_{\rm core}}{\hat{H}_c}
  -\frac{(N_1-N_2)}{(N_2+1)}
\right\}^{N_2}.
\eeq

Note that upon setting $N_1=N_2$, both of the above relations reduce
to the single-polytrope case of equation \eqref{iterate_poly}, as required.

\subsection{Iterative solution: the magnetic field}

As mentioned above, for magnetised stellar models the extended HSCF
method also requires us to be able to find an integral equation
incorporating information about the star's magnetic field. The
description here is breviloquent, since detailed
derivations may be found elsewhere (e.g. \citet{LJ09}).

If we assume the star to be axisymmetric and work in
cylindrical polar coordinates $(\varpi,\phi,z)$ aligned so that the $z$ coordinate is
the star's symmetry axis, then one can show from the constraint
$\div\bB=0$ that the magnetic field can be expressed in the form
\beq
\bB=\bB_{\rm pol}+\bB_{\rm tor}
      =\frac{1}{\varpi}\brac{\nabla u\times\be_\phi+f(u)\be_\phi},
\eeq
where $u$ is the poloidal magnetic streamfunction, defined through this
expression, and $f(u)$ is a function of $u$ -- which from a
mathematical perspective is virtually arbitrary, but physically
relates to the toroidal-field component. In order to avoid toroidal
field -- and therefore an electric current -- outside the
star, the function $f$ needs to be fitted to a contour of $u$ which
closes within the star. If we define $u_{\rm max}$ as the largest such
contour (i.e. the last field line which closes inside the star), then
\beq
f(u)=a(u-u_{\rm max})^{\zeta}\mathcal{H}(u-u_{\rm max}),\zeta>1,
\eeq
where $\mathcal{H}$ is the Heaviside function and $a$ and $\zeta$ are constants.
It is clear from the form of the Lorentz force that $\nabla
M\cdot\bB=0$, and from the expression for $\bB$ in terms of $u$ we
also have $\nabla u\cdot\bB=0$. The two gradients $\nabla M$ and
$\nabla u$ are therefore parallel, and so we deduce that
\beq
M=M(u).
\eeq
One can then derive a single differential equation in the variable
$u$, which -- together with the chosen prescriptions for $M(u)$ and
$f(u)$ -- encodes all the information about the magnetic field. This
is known as the Grad-Shafranov equation, and has the form:
\beq\label{GS}
\Delta_* u = -4\pi\rho\varpi^2\td{M}{u}-f\td{f}{u},
\eeq
where the differential operator $\Delta_*$ is the axisymmetric Laplacian operator,
but with the opposite sign on the first-derivative piece:
\beq
\Delta_*\equiv \pd{^2}{\varpi^2}-\frac{1}{\varpi}\pd{}{\varpi}+\pd{^2}{z^2}.
\eeq
Exploiting a standard Green's function, equation \eqref{GS} may be
written in a Poisson-like integral form \citep{tomi_eri,LJ09}, completing the system of
integral equations. It needs to be solved to find $u$ at each
iterative step, using the $u$ and $\rho$ distributions from the
previous step. With the updated solution for $u$, one then evaluates $M(u)$
with it, and uses this in the first integral of the Euler equation. In
this way, the magnetic field and the density distribution are
self-consistently updated at each iterative step:\\
(i)  we account for the effect of the density 
  distribution and from the different forces in the star on the
  magnetic field;\\
(ii) we account for the distortion to the density distribution
  induced by the magnetic field.

\subsubsection{Function choices in this paper}
\label{func_choice}

Other than the restrictions described above, the functional forms of $f(u)$ and
$M(u)$ may be chosen freely, although varying these has limited effect
on the resulting equilibria, if they are found using a self-consistent
method; see \citet{LJ12} or \citet{bucc15} for a survey of these parameters. The
constant $\kappa$ sets the overall field strength, and $a$ the maximum
strength of the toroidal component; the value of $\zeta$ is less
important. For all
poloidal/linked poloidal-toroidal field
results in this paper we take $\zeta=0.01$ and
$M(u)=\kappa (u/u_{\rm gmax})^5$ (where $u_{\rm gmax}=\max(u)$);
we have found that these allow for the maximum 
strength of toroidal field in our linked poloidal-toroidal
magnetic-field solutions (which are always dominated, energetically,
by the poloidal component). For purely toroidal fields -- a different
class of solution where there is no additional equation like equation
\eqref{GS} to solve --  we take
$M=-\lambda^2\rho\varpi^2/4\pi$, where $\lambda$ is a constant
governing the field strength \citep{LJ09}.

\subsection{Physical sequences of models}

In cold polytropic models of NSs, one calculates a single dimensionless model
(the most natural choice being an unmagnetised non-rotating one), chooses the
desired physical mass $\mathcal{M}$, and finds the value of the (single)
polytropic constant $k$ that gives the desired physical radius $R_*$. Any
two models with the same physical $\mathcal{M},k$ can be regarded as the same
physical star; we therefore restore the dimensions of other models
(rotating and/or magnetised), by multiplying by the requisite combination of
$\rho_c$ and $R_*$ (found from the fixed physical $\mathcal{M},k$ and
the dimensionless $\hat{\mathcal{M}},\hat{k}$ calculated for an
individual model).

Here, with a two-piece polytrope and hot models,
the procedure is less general but similar. We again fix a
non-rotating, unmagnetised, and now also zero-$T$ model; in all
results reported here this spherical reference model has
$\mathcal{M}=1.4\mathcal{M}_\odot$, where $\mathcal{M}_\odot$ is the mass of the Sun, and $R_*=12$ km. We run the code for such a
model and obtain the dimensionless polytropic constant $\hat{k}_2$ for
the core from equation \eqref{Hc} and the dimensionless mass
$\hat{\mathcal{M}}$ by volume integration of $\hat\rho$. Now, since
these two dimensionless quantities are related to their physical
counterparts by:
\beq
\hat{\mathcal{M}}=\frac{\mathcal{M}}{\rho_c R_*^3},\
\hat{k}_2=\frac{k_2}{G\rho_c^{1-1/N_2}R_*^2},
\eeq
we may combine these relations to calculate the physical value of
$k_2$ for the reference model:
\beq\label{k2_phys}
k_2=\hat{k}_2 G\brac{\frac{\mathcal{M}}{\hat{\mathcal{M}}}}^{1-1/N_2} R_*^{3/N_2-1}.
\eeq
Recall that the envelope polytropic constant $k_1$ is not
independent of $k_2$, and hence the corresponding relation for $k_1$
gives no additional information. We choose to work with $k_2$, as the core
comprises most of the mass and volume of the star. A physical sequence
of models, therefore, has fixed $k_2$ and
$\mathcal{M}$ in physical units. Their dimensionless
counterparts will, however, vary from model to model depending on the
star's rotation rate, magnetic field
and temperature. Using these physical and dimensionless quantities, we
are now able to calculate the physical equatorial radius for any given
model, through a rearrangement of equation \eqref{k2_phys}:
\beq
R_*=\left[\hat{k}_2 G\brac{\frac{\mathcal{M}}{\hat{\mathcal{M}}}}^{1-1/N_2}\right]^{N_2/(N_2-3)}.
\eeq
Having done so, we are then able to calculate the central density in
physical units:
\beq
\rho_c=\frac{\mathcal{M}}{\hat{\mathcal{M}}R_*^3}.
\eeq
Since $\rho_c$ enters the iterative procedure for both hot and
piecewise-polytropic models, we must recalculate $R_*$ and $\rho_c$
using the above relations at each iterative step.

\subsection{Iterative scheme}

The HSCF-based numerical scheme we use iterates towards a solution by
using the equilibrium equations in integral form. The scheme takes the form:\\
0. As initial conditions to start an iteration, make simple trial guesses for
$\rho$ and $u$;\\
1. Calculate the gravitational potential $\Phi$ from the $\rho$
distribution and Poisson's equation \eqref{poisson} in integral form;\\
2. Calculate the new magnetic streamfunction $u$ from its
previous form $u_\textrm{old}$, using the magnetic
Poisson equation \eqref{GS} (in integral form) with $u_\textrm{old}$ and $\rho$ in the
integrand;\\
3. Calculate, in physical units, $R_*$ and $\rho_c$, and use these to
calculate the thermal-force scalar $\hat\Theta$;\\
4. Evaluating the Euler equation at the equatorial
and polar surfaces, equations \eqref{Om2} and \eqref{intC}, find $\hat\Omega^2$ and $\hat{C}$;\\
5. We are now able to use the Euler equation \eqref{int_Euler_hot} to find $H$
throughout the star;\\
6. Calculate the new density distribution in the envelope and core
with equations \eqref{iter_rho_env}, \eqref{iter_rho_core};\\
7. For stability reasons we do not always use the fully-updated $u,\rho$
distributions for the following iterative step, but instead employ an
underrelaxation step. We then return to step 1 using the
partially-updated $\rho$ and $u$ distributions, repeating the cycle
until satisfactory convergence is achieved, i.e. until the fractional changes in
$\hat{H},\hat{C},\hat\Omega^2,\hat{u}$ between consecutive iterative 
steps drop below some small tolerance value (usually $10^{-4}-10^{-5}$).\\

The input parameters for any equilibrium configuration are the surface
distortion $r_\textrm{pole}/r_{\textrm{eq}}$, the polytropic
indices in the core and envelope regions $N_1,N_2$ and prefactors $a,\kappa$
related to the strengths of the poloidal and toroidal field
components. In the purely-toroidal case there is a single constant
$\lambda$ to specify. The grid is evenly-spaced in $r$ and
$\cos\theta$; the latter ensures that the equatorial region is well resolved even with a limited
number of angular grid points. This is important since this region can have
complex field geometry, with coexisting poloidal and toroidal
components, and strong variations in the density for models rotating
close to Keplerian velocity -- whereas the polar region is relatively
featureless. Since the density and physics of the
envelope region changes over a radius $\sim 0.1R_*$, good coverage of
this region is also needed. For these reasons we have found a good
grid resolution, which we adopt as our standard here, consists of $512$
radial gridpoints and $128$ angular gridpoints.

The code exploits a decomposition of the governing equations into
multipoles.  Since the models are axisymmetric the azimuthal index $m$
is zero, and the equations become an infinite expansion in terms of
Legendre polynomials with angular index $l$. For numerical purposes
this clearly must be terminated at some maximum $l=l_{\rm max}$, at
which the contribution of additional multipoles should be
negligible. For more extreme models -- a very strong toroidal
component or very rapid rotation -- we have found that very high
multipoles can make a visible difference to the final magnetic-field
configuration (see section \ref{multipoles}), and so we choose $l_{\rm max}=32$ as standard in this
work.

The iterative process described here typically takes of the order
$10-500$ steps, and even for high resolutions finishes within a few
minutes when run on a typical laptop. The code is stable up to $s_{b0}=2$
in many cases, and up to $s_{b0}=1.5$ for extremal models (i.e. Keplerian
rotation and/or strong magnetic fields); this is certainly adequate,
since higher values of entropy are not consistent with our hot EOS model anyway.

\section{Results}
\label{results}

\subsection{Virial test}

\begin{figure} 
\begin{center}
\begin{minipage}[c]{\linewidth}
  \psfrag{NDIV}{$N_{\rm DIV}$}
  \psfrag{VT}{$VT$}
\includegraphics[width=\linewidth]{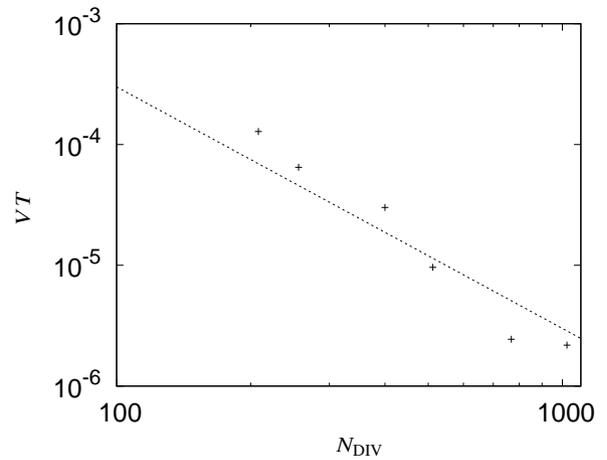}
\caption{\label{virial}
  Convergence of code accuracy with increased resolution, for a hot,
  rapidly-rotating and highly magnetised model ($s_{b0}=1.0$,
  $\Omega=690$ Hz, $\hat\kappa=0.3,\ \hat{a}=12$).
  Points show virial-test results $VT$ for different numbers of radial
  grid points $N_{\rm DIV}$, and the dashed line shows the expected
  behaviour for a second-order code, of inverse-square scaling with
  grid resolution.}
\end{minipage}
\end{center}
\end{figure}

Before moving to our results, we first confirm that our numerical code
is behaving as expected. The natural measure of the accuracy
of such a code comes from the virial theorem, in which the vector Euler equation is
converted into a scalar energy balance:
\beq
\mathcal{E}_{\rm grav}+3\Pi_0+3\Pi_{\rm th}+2\mathcal{E}_{\rm kin}+\mathcal{E}_{\rm mag}=0,
\eeq
where $\mathcal{E}_{\rm grav}, \mathcal{E}_{\rm kin}, \mathcal{E}_{\rm mag}$
are the gravitational binding, kinetic and magnetic energies; and
$\Pi_0,\Pi_{\rm th}$ the volume integrals of the zero-temperature and
thermal pressures. In the above form of the virial theorem
the right-hand side is zero, reflecting the fact that the
solution should be a stationary equilibrium. The left-hand side is evaluated for the
solutions produced by the numerical scheme, and then normalised by
dividing by $|\mathcal{E}_{\rm grav}|$, to give a dimensionless
measure of the code's accuracy: the virial test $VT$. In figure
\ref{virial} we present values of $VT$ for a numerically challenging model to
calculate (hot, highly-magnetised and rapidly-rotating) as a function
of grid resolution. We confirm that the error is very small compared
with unity, and furthermore that it drops with increasing resolution in the manner expected
for a second-order convergent scheme (the order at which the code is
written). For all results presented in this paper the virial test has
also been checked; it is never more than order $10^{-4}$, and in many
cases is as low as $10^{-6}$, comparing favourably with other studies.

\subsection{Keplerian velocity}

With progressively more rapid rotation, a star becomes more oblate,
until -- at some critical rotation rate -- it begins to lose mass from
the equatorial surface. At this \emph{Keplerian} rotation rate
$\Omega_K$ the centrifugal force at the equatorial surface matches the
gravitational force, $\nabla\Phi_r=\nabla\Phi$. To check when this is
reached, we evaluate the auxiliary quantity
\beq
\Omega_c^2=\frac{1}{R_*}\pd{\Phi}{r}.
\eeq
This quantity is generally less than $\Omega^2$, but when Keplerian
velocity is reached the two become equal:
$\Omega_c=\Omega=\Omega_K$. Models with $\Omega>\Omega_c$ are
unphysical for our purposes, as the star would be in a dynamical
mass-shedding state. Clearly we cannot calculate a model that
precisely satisfies the Keplerian condition; instead, we repeatedly
run the code to find the equilibrium model where $\Omega$ is the
closest to (but still less than) $\Omega_c$. The rotation rate of this
model is then recorded as $\Omega_K$.  Therefore, all results for
$\Omega_K$ are very slight underestimates.

\subsection{Hot unmagnetised models}

We begin by exploring the stellar structure of our proto-NS models and
comparing with their zero-temperature counterparts. To study the
effect of rotation on hot and cold NSs, we look at the two extremes of
non- and maximally-rotating NSs (i.e. those rotating at
Keplerian velocity). We have also checked the corresponding results
for magnetised stars, finding that none of the results reported here
are modified unless the magnetic field is substantially stronger than
$10^{16}$ G -- and since there is no good physical reason to expect
such strong fields in newborn NSs, we do not consider this case
further. In addition, although we show results only for the piecewise
polytrope with $N_1=4,N_2=0.6$, we have also run many models for the
case $N_1=3,N_2=0.6$ and some other variations, finding no significant
differences in the results.

\begin{figure} 
\begin{center}
\begin{minipage}[c]{0.9\linewidth}
\psfrag{Tc_11}{$T_c$ [$10^{11}$ K]}
\psfrag{sb0}{$s_{b0}$}
\includegraphics[width=\linewidth]{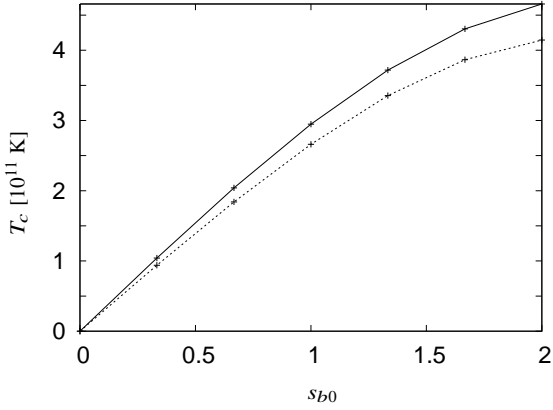}
\caption{\label{Tcent_vs_s}
  Central temperature as a function of $s_{b0}$ for non-rotating models
  (solid line) and their counterparts at Keplerian velocity (dashed line).}
\end{minipage}
\end{center}
\end{figure}

\begin{figure*} 
\begin{center}
\begin{minipage}[c]{\linewidth}
  \psfrag{sb}{$s_b$}
  \psfrag{Temp}{$T$ [K]}
  \psfrag{Theta}{$\hat\Theta$}
  \psfrag{r}{$r/R_*$}
\includegraphics[width=\linewidth]{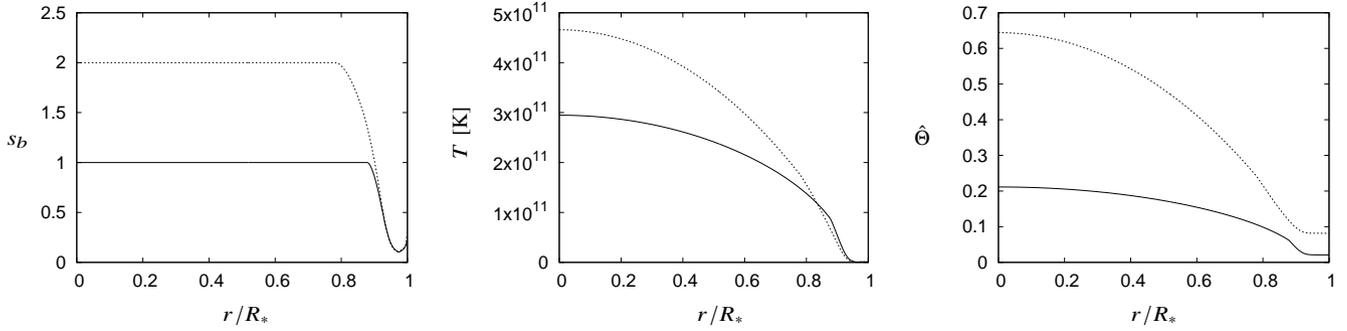}
\caption{\label{therm_quant}
  Radial profiles of entropy per baryon $s_b$, temperature and the
  dimensionless thermal-force scalar for two non-rotating unmagnetised
  models with $s_{b0}=1$ (solid line) and $s_{b0}=2$ (dashed
  line). Note that the physical radii of the two models are different;
plotting against dimensionless radius allows for an easier comparison
of features.}
\end{minipage}
\end{center}
\end{figure*}

In our models, the fundamental parameter determining the importance of
thermal effects is the central entropy $s_{b0}$, but it is often more
useful to know the star's temperature. For this reason we begin our
survey of models by comparing central temperature and entropy; see
Fig. \ref{Tcent_vs_s}. The relationship is little affected by
rotation, with the lines for $\Omega=0$ and $\Omega=\Omega_K$ very
close to one another; the non-rotating results are well fitted by the
following relation:
\beq\label{T-s_approx}
T_c[10^{11}\textrm{ K}]=-0.58s_{b0}^2+3.53s_{b0}.
\eeq

Fig. \ref{Tcent_vs_s} is complemented by Fig. \ref{therm_quant}, which
shows the radial profiles of the fundamental thermal quantities: the
entropy density, temperature and thermal-pressure scalar. The
smoothness of these quantities across the envelope-core and
inner-outer envelope boundaries -- where the physics of the star
changes -- vindicates our prescription for the thermal physics.

\begin{figure} 
\begin{center}
\begin{minipage}[c]{0.9\linewidth}
\psfrag{T}{$T$ [MeV]}
\psfrag{sb}{$s_b$}
\psfrag{encmass}{$\mathfrak{m}\ [\mathcal{M}_\odot]$}
\includegraphics[width=\linewidth]{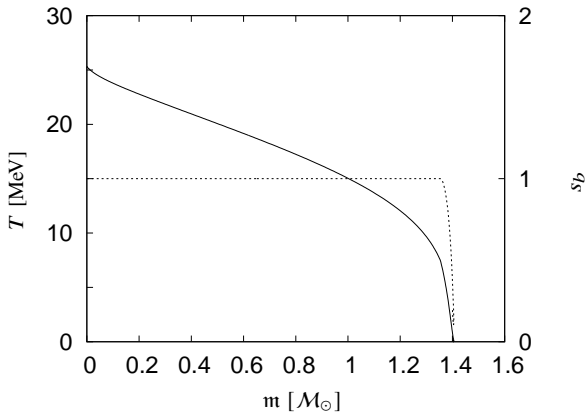}
\caption{\label{sT_encm}
  Temperature $T$ (solid line) and entropy $s_b$, both given in
  typical nuclear-physics units, as a function of enclosed mass $\mathfrak{m}$
  in units of $\mathcal{M}_\odot$, to allow for direct comparison with
  plots from the proto-NS literature. The model shown is non-rotating
  and unmagnetised, with $s_{b0}=1$.}
\end{minipage}
\end{center}
\end{figure}

Next we compare our temperature and entropy profiles with
detailed quasi-equilibrium calculations for non-rotating proto-NSs by
\citet{BL86} and \citet{pons99}, hereafter BL86 and P99. Although we cannot expect exact
agreement given our simplified model, our results should at least be
sensible. In Fig. \ref{sT_encm} we replot the $s_b$ and $T$ profiles
for the $s_{b0}=1$ model from Fig. \ref{therm_quant}, but as a
function of enclosed mass $\mathfrak{m}$ rather than radius, and with
$T$ in MeV. We compare with figures 1 and 2 of BL86, whose
fiducial model is $1.4\mathcal{M}_\odot$ like ours, and figure 9 of
P99, for a $1.6\mathcal{M}_\odot$ model -- in all cases,
looking at results after several seconds, when the shocked mantle has
cooled and the temperature is highest at (or very close to) the centre of
the star.

We confirm that our isentropic
assumption was not heinous: in the realistic profiles $s_b$ never
varies by more than a factor of $\sim 2$ for the latter phase of the
proto-NS evolutions. The entropy reaches an average value of roughly
unity at a time of $15$ s for the BL86 simulation, and $30$ s
for that of P99, so let us compare the corresponding $T$
profiles with ours for $s_{b0}=1$. The central temperature for our
model is $25$ MeV, close to both P99 (also $\sim 25$ MeV) and BL86
($\sim 20$ MeV). The $T$ profiles of BL86 and P99 both decrease by a
factor of $\sim 5$ before a rapid drop in the outermost region
(presumably the envelope). Our $T$ profile shows the same kind of
behaviour, but with a gentler drop over the core region: at the
boundary with the envelope the temperature is a factor of $3.4$
smaller than in the centre. These differences are relatively minor,
considering that we do not treat any of the important neutrino physics
and neglect the factor-$2$ variation of $s_b$ within the star, and so
we conclude that our model is a sensible approximation to the full problem.

\begin{figure} 
\begin{center}
\begin{minipage}[c]{\linewidth}
\psfrag{logrho}{$\log(\rho[\textrm{g cm}^{-3}])$]}
\psfrag{rad}{$r/R_{\rm eq}$}
\includegraphics[width=\linewidth]{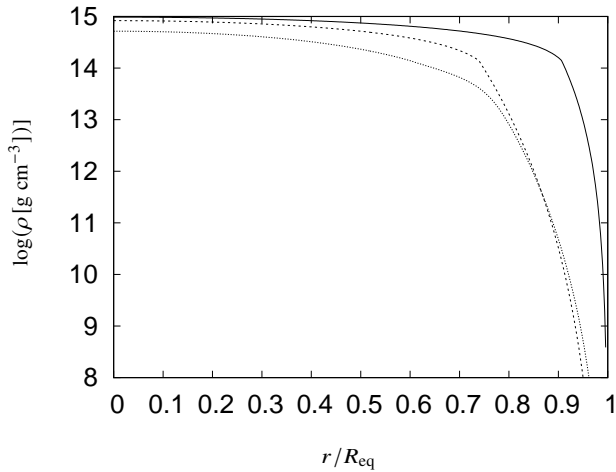}
\caption{\label{rho_prof}
  Density profiles along the equator ($x$-axis) as a function of
  dimensionless radius. Solid, dashed and dotted lines correspond to:
  $s_{b0}=0,\Omega=0$ ; $s_{b0}=0,\Omega=\Omega_K$ ; $s_{b0}=2,\Omega=\Omega_K$. }
\end{minipage}
\end{center}
\end{figure}

Next we study the physics of proto-NSs rotating at Keplerian velocity
through a series of figures. First, in Fig. \ref{rho_prof}, we compare
the equatorial density profiles of three model stars. The actual radii
differ for each star, but they are plotted together using the
normalised radius $\hat{r}=r/R_*$ for direct comparison. The profile for
the cold, non-rotating model shows the expected shape for a mature
neutron star: a core region extending to a radius $r\sim 0.9R_*$, with
density decreasing by only a factor of a few, followed by a plunge of
the density towards zero over the last $\sim 0.1R_*$ of the star's
radius. In comparison with this, the same cold model rotating at
Keplerian velocity has a more extended envelope, covering the
equatorial radius $r\gtrsim 0.75 R_*$, with $\rho$ again descending
smoothly to zero at the stellar surface. Finally, we compare this
maximally-rotating cold model with an extremely hot counterpart. The
hot model also has an extended envelope, but with a smoother
transition at the envelope-core boundary. In the hot envelope
$\rho$ descends more gradually than in the cold model, being held up
by the thermal pressure.

\begin{figure*} 
\begin{center}
\begin{minipage}[c]{\linewidth}
\includegraphics[width=\linewidth]{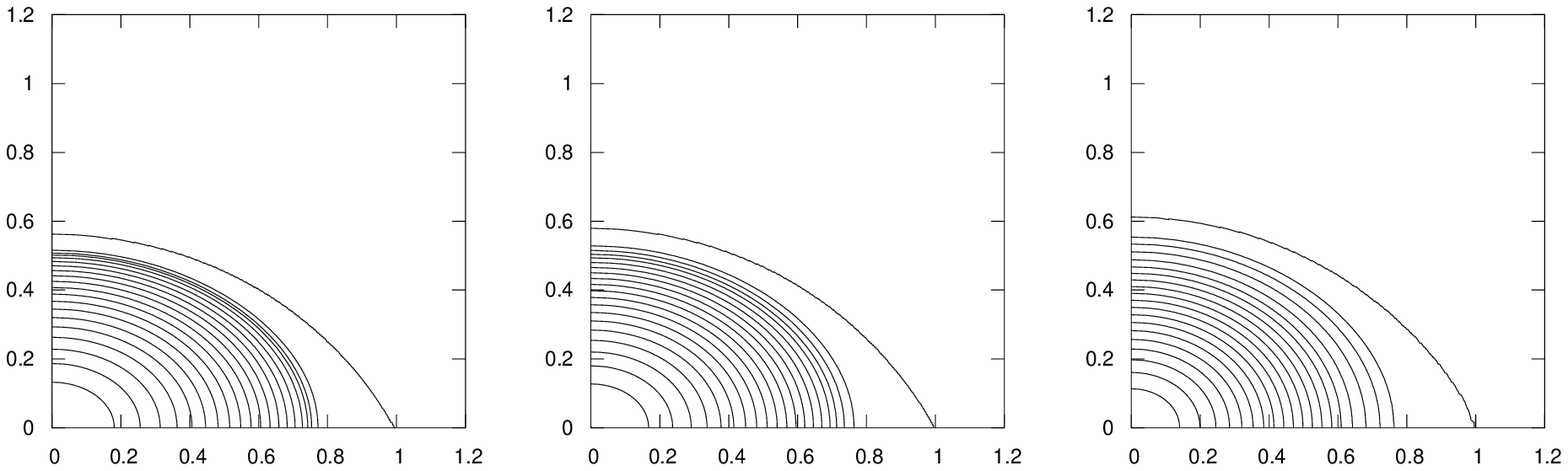}
\caption{\label{dens-kep}
  Density contours of three NS models rotating near Keplerian
  velocity; the outer contour shows the stellar surface. Units are
  dimensionless radii, normalised to the equatorial stellar surface. From left to right, $s_{b0}=0,1,2$.}
\end{minipage}
\end{center}
\end{figure*}

\begin{figure*} 
\begin{center}
\begin{minipage}[c]{0.8\linewidth}
\psfrag{rhoc}{$\rho_c$ [$10^{14}$ g cm${}^{-3}$]}
\psfrag{rad}{$R_*$ [km]}
\psfrag{sb0}{$s_{b0}$}
\includegraphics[width=\linewidth]{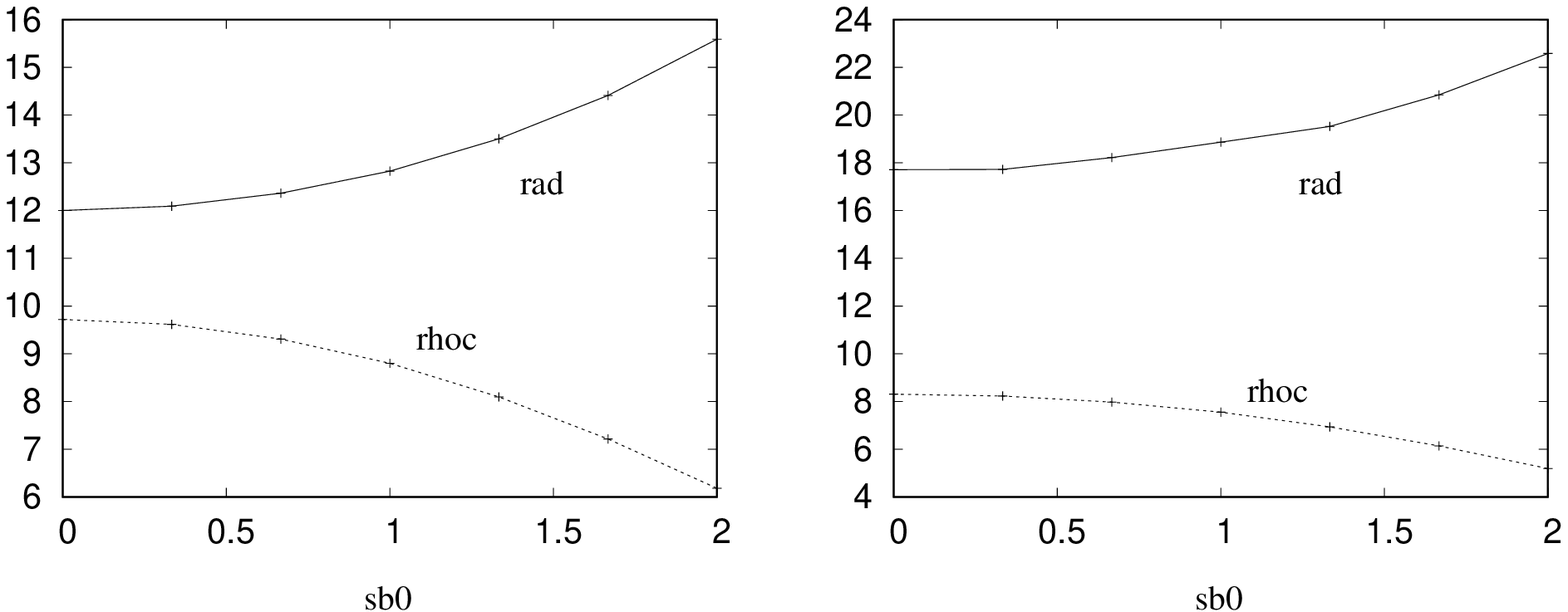}
\caption{\label{radrho_s}
  Dependence of $\rho_c$ and $R_*$ on central entropy per unit baryon
  $s_{b0}$ (in $k_B$) for a set of models with mass $1.4\mathcal{M}_\odot$ and
  the same polytropic constants and indices. Left: non-rotating models, right: models
  rotating at Keplerian velocity (here $R_*$ is equatorial
  radius).}
\end{minipage}
\end{center}
\end{figure*}

We have seen the effect of Keplerian velocity and high temperature
along an equatorial radial spoke; we now look at the rest of the
star's density distribution, through the contour plots of
Fig. \ref{dens-kep}. With twenty equally-spaced contours in each case,
we see a bunching of contours in the outer core followed by a single
extended region, wider at the equator, corresponding to the
envelope. The contours are slightly smoothed out at higher
temperatures. What the plot cannot convey is the changes in central
density and radius, so we plot the variation of these with $s_{b0}$ in
Fig. \ref{radrho_s}. We see the equatorial radius of our canonical model -- 12 km
at zero temperature and without rotation -- can almost double for the
hottest model at Keplerian rotation. At the same time, the central
density roughly halves.

\begin{figure} 
\begin{center}
\begin{minipage}[c]{\linewidth}
\psfrag{OmK}{$\nu_K$ [Hz]}
\psfrag{sb0}{$s_{b0}$}
\includegraphics[width=\linewidth]{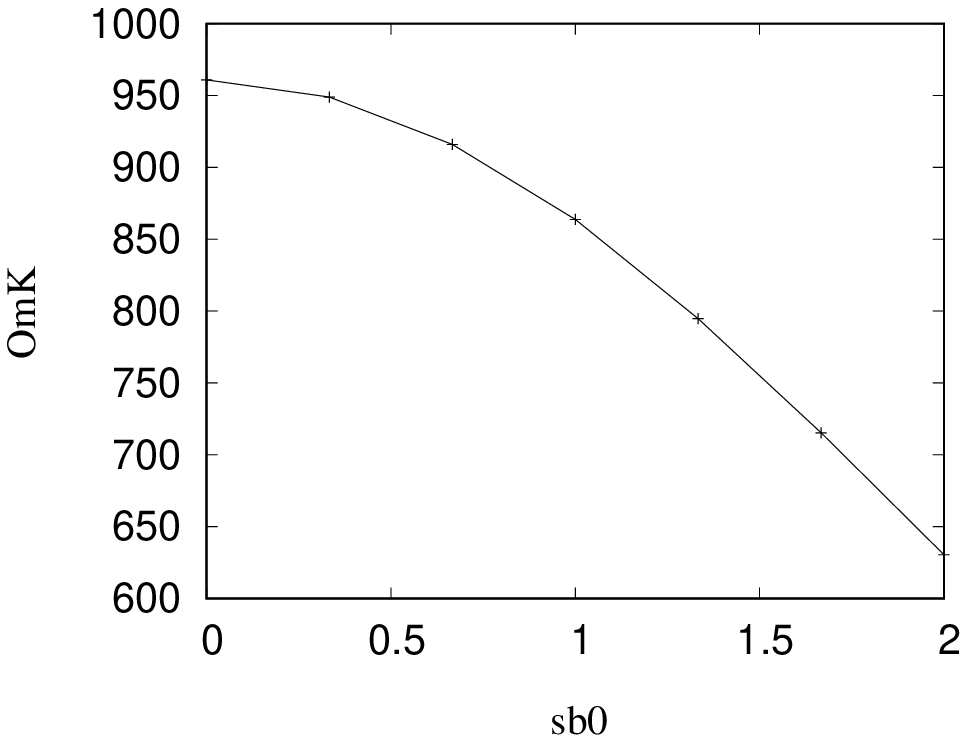}
\caption{\label{kep_vs_s}
  Keplerian rotational frequency $\nu_K=\Omega_K/2\pi$ as a function of $s_{b0}$. Very hot
  models are seen to break up at notably lower rotation rates.}
\end{minipage}
\end{center}
\end{figure}

Finally, we plot the effect of increasing temperature on the Keplerian
rotation rate of the star in Fig. \ref{kep_vs_s}. For the hottest
model this maximum
rotation rate decreases rather dramatically, by roughly one third,
compared with the cold model. Recall that we have checked this
behaviour is not peculiar to our particular choice of envelope
polytropic index $N_1=4$, but is seen with lower values of $N_1$ too.
The results for Keplerian configurations plotted in Figs. \ref{dens-kep}-\ref{kep_vs_s} are in
very good agreement with the work of \citet{haensel09}, who present
approximate relations for stars at Keplerian rotation as a function of
their non-rotating counterparts. In particular, with their formula
$R_*(\mathcal{M},\nu=\nu_K)=1.44 R_*(\mathcal{M},\nu=0)$
one can accurately
predict the radii in the right-hand panel of Fig. \ref{radrho_s} given
the values of the left-hand panel. Another formula,
$\nu_{\rm K}=1.08\,{\rm kHz} (\mathcal{M}/\mathcal{M}_\odot)^{1/2}(R/10\,{\rm km})^{-3/2}$,
successfully reproduces Fig. \ref{kep_vs_s}, again given the
$\nu=0$ values for equatorial radii (recall that all our models
presented here have mass $\mathcal{M}=1.4\mathcal{M}_\odot$).

\subsection{Magnetic-field structures}
\label{mag_struct}

\begin{figure*}
\begin{center}
\begin{minipage}[c]{0.7\linewidth}
\includegraphics[width=\linewidth]{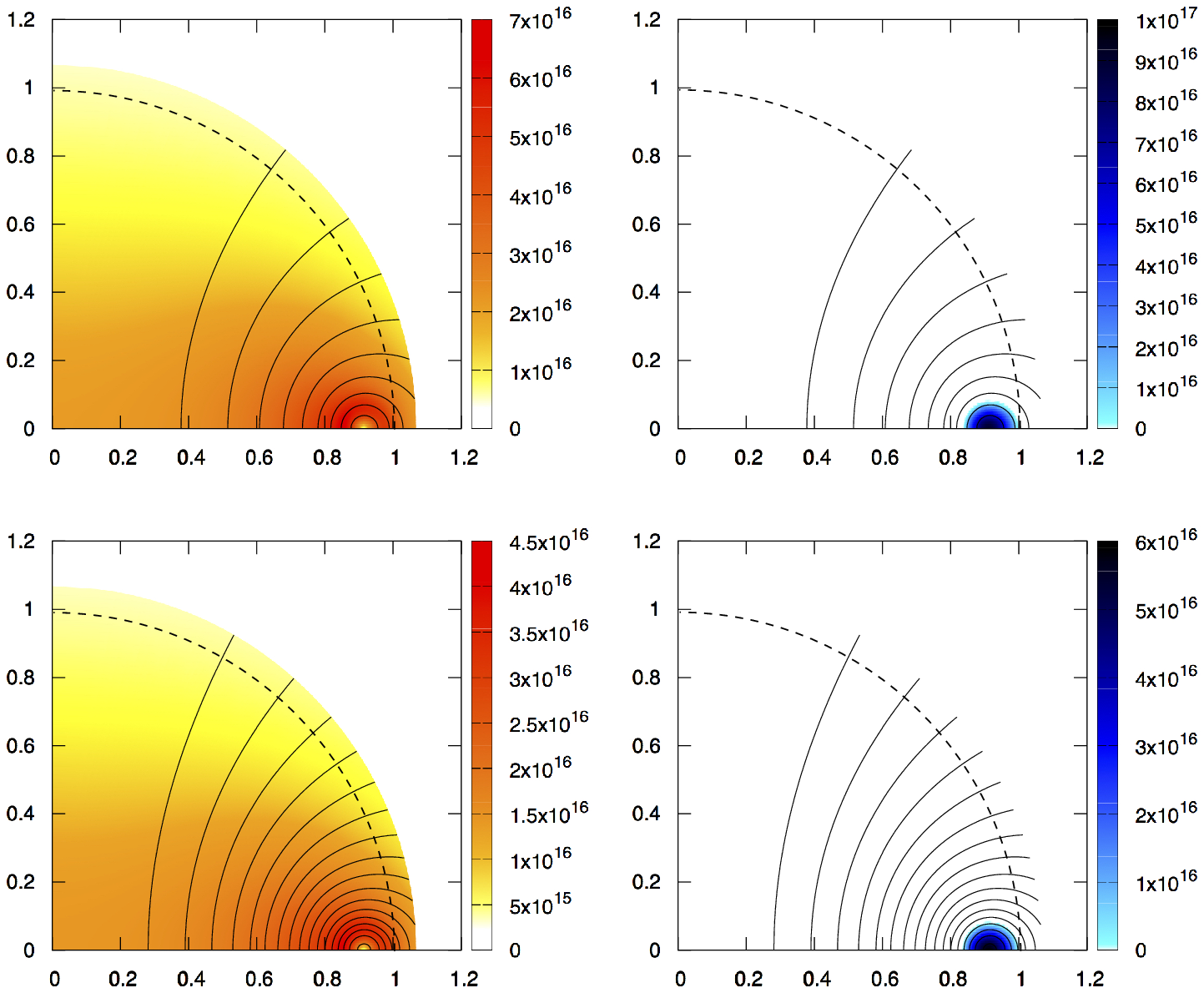}
\caption{\label{mix_field}
  Two non-rotating models, slightly oblately distorted by the magnetic
  field. With the colourscales we plot the magnitude of the poloidal (left) and toroidal (right) field
  components, in units of gauss, together with the poloidal field
  lines. The dashed lines show the stellar surface. Top: $s_{b0}=0$,
  bottom: $s_{b0}=1.5$. These models
  feature around the strongest toroidal component, in terms of its
  contribution to the total magnetic energy, that our numerical
  method is able to find: $\mathcal{E}_{\rm mag}^{\rm tor}/\mathcal{E}_{\rm mag}= 8.5\%$ for
  both. The magnetic-field structure of the hot model is virtually
  identical to the cold one, although the magnitudes of the field
  components are lower.}
\end{minipage}
\end{center}
\end{figure*}

We now present some representative results for the magnetic field of a
late proto-NS.  Firstly, we look at a linked poloidal-toroidal
magnetic field configuration; see Fig. \ref{mix_field}. A very hot
model, with $s_{b0}=1.5$, is compared with its zero-temperature
counterpart. Although non-rotating, the two stars are slightly oblate by
virtue of dominantly-poloidal magnetic fields (strong toroidal
fields, by contrast, induce prolate distortions). In fact, all such
self-consistent zero-temperature equilibria found to date feature a
poloidal component that is energetically dominant, with the magnetic
energy in the toroidal component $\mathcal{E}_{\rm mag}^{\rm tor}$
being only a small fraction of the total; one motivation for
the work reported here was to see whether the same remained true for
hot proto-NSs.

Fig. \ref{mix_field} demonstrates that the temperature of a NS plays
essentially no role in determining the star's magnetic-field
structure, with the two models being indistinguishable. This strongly
suggests that our simplified model for the thermal physics is
perfectly adequate for this problem. Although we have chosen free
functions in order to maximise
the importance of the toroidal component (see section \ref{func_choice}), only $8.5\%$
of the magnetic energy is stored in the toroidal component in both
the hot and the cold models.
The key difference is in the magnitude of the magnetic field, showing
that a hot NS is more readily distorted by a magnetic field than its
cold counterpart. We will explore this more in section \ref{ellip}.

\begin{figure*}
\begin{center}
\begin{minipage}[c]{\linewidth}
\includegraphics[width=\linewidth]{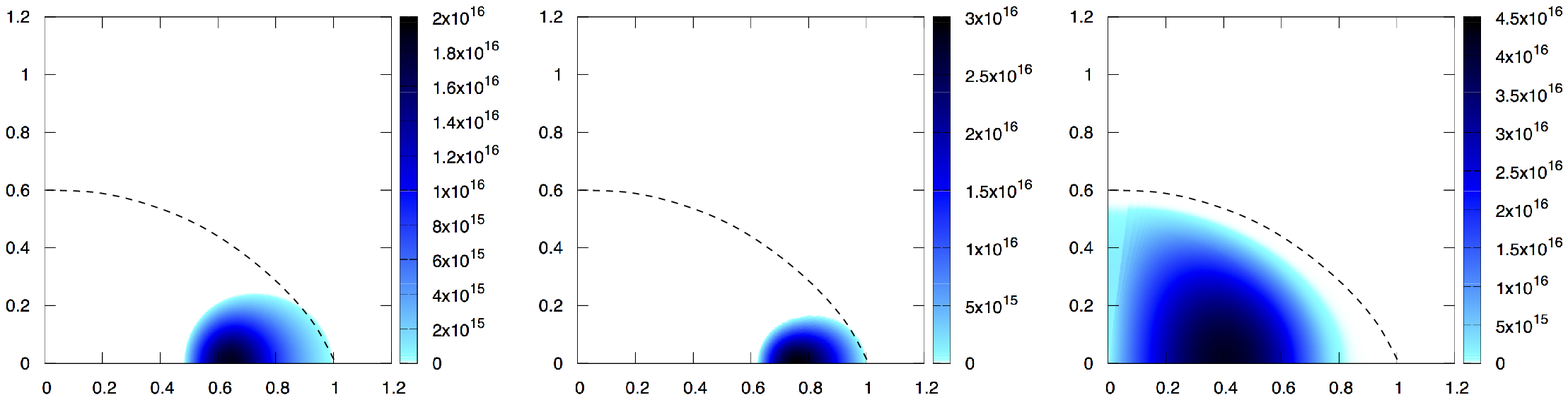}
\caption{\label{Btor_Kep}
  Toroidal field strength (colourscale) for three models with
  $s_{b0}=1.5$ and $\Omega=\Omega_K$, and all with an average internal field strength of
  $2\times 10^{16}$ G. The dashed line denotes the stellar surface. Left and middle:
  the toroidal component of a linked poloidal-toroidal field model
  with $\mathcal{E}_{\rm mag}^{\rm tor}/\mathcal{E}_{\rm mag}= 5\%$ (left) and $10.7\%$
  (middle). Right: a purely-toroidal field model.}
\end{minipage}
\end{center}
\end{figure*}

Finally, in Fig. \ref{Btor_Kep} we compare the distribution of
toroidal field within two linked poloidal-toroidal models, and one
purely-toroidal model, all rotating at Keplerian velocity. We consider
three hot proto-NSs models, with $s_{b0}=1.5$; again, their cold
counterparts are very similar in structure, but with different
magnitudes. In the linked poloidal-toroidal models, we see that a
slightly stronger toroidal component is possible compared with the
almost-spherical non-rotating models of the previous figure: one model
has  $\mathcal{E}_{\rm mag}^{\rm tor}/\mathcal{E}_{\rm mag}= 10.7\%$. We see an effect
already known from cold models \citep{bucc15,armaza}: as the maximum
strength of the toroidal component is increased, the region it
occupies decreases, leading to locally-intense toroidal fields whose
contribution to the total magnetic energy is no larger than for
locally-weaker counterparts.

\subsection{Ellipticity}
\label{ellip}

The magnetically-induced ellipticity, measuring the distortion from
sphericity of a star's mass distribution, is of interest, as a star
with misaligned rotation and magnetic axes will emit continuous GWs at
a magnitude proportional to this distortion \citep{bon_gour}. We find that it is
somewhat easier to distort a hot star than a cold one. We constructed
a number of magnetised and non-rotating models for a given $s_{b0}$,
always finding that the results were in excellent agreement with the expected
quadratic scaling $\epsilon=kB^2$. We then repeated the procedure for
different values of $s_{b0}\leq 1.5$, finding that increases in
$\epsilon$ were proportional to $s_{b0}^2$. Combining these results, we find a reasonable fit
(deviating by less than 3\% from all results) to the magnetically-induced
ellipticity of a hot NS to be:
\beq
\epsilon=10^{-5}(3.0s_{b0}^2+8.3)
                   \brac{\frac{B_{\rm pole}}{10^{15}\textrm{ G}}}^2
          =10^{-6}(2.3s_{b0}^2+6.5)
                   \brac{\frac{B_{\rm av}}{10^{15}\textrm{ G}}}^2
\eeq
for poloidal fields, and
\beq
\epsilon=-10^{-6}(2.1s_{b0}^2+6.9)
                   \brac{\frac{B_{\rm av}}{10^{15}\textrm{ G}}}^2
\eeq
for toroidal fields. In this latter case the ellipticity is negative,
since the induced distortion is prolate. The results are only reported
as a function of $B_{\rm av}$, since the toroidal magnetic-field strength drops
to zero at the surface. The above formulae can readily be converted to
a function of central temperature instead of entropy, using equation
\eqref{T-s_approx}.
Finally, results for mixed poloidal-toroidal
models are not reported; the toroidal component only marginally reduces the
oblateness (and therefore the ellipticity), since it occupies an
insignificant low-density fraction of the stellar volume.

In the quadrupole formula for gravitational radiation, the ellipticity multiplies the moment of
inertia of the star. The increase in $\epsilon$ we find could conceivably have
been cancelled by a corresponding decrease in the moment of inertia,
thus leading to no enhancement in the GW signal; however, upon
checking this we found the moment of inertia varies very little with
temperature (and, in fact, increases slightly).

\subsection{Multipolar structure}
\label{multipoles}

\begin{figure*}
\begin{center}
\begin{minipage}[c]{\linewidth}
\includegraphics[width=\linewidth]{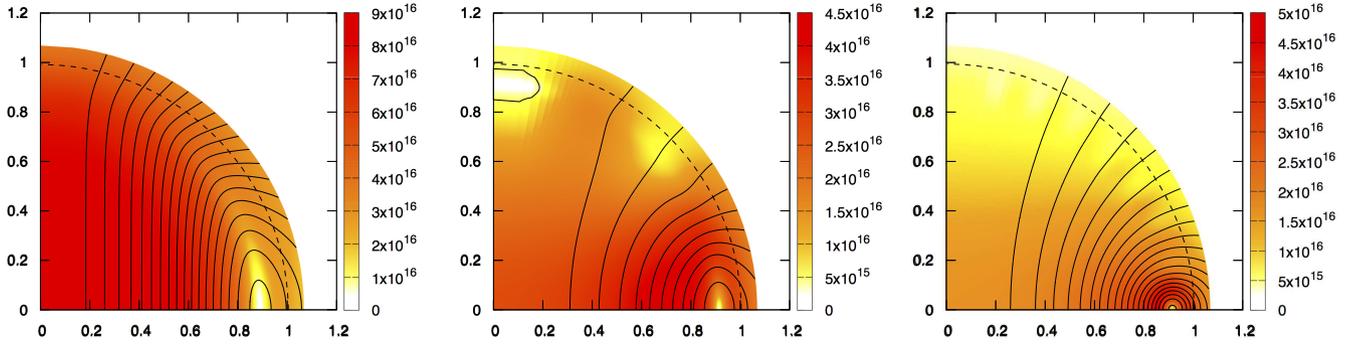}
\caption{\label{multipole_plot}
  The effect of truncating equilibrium solutions at different
  multipoles. The magnitude (colourscale) and direction (lines) of the
poloidal component of a linked poloidal-toroidal field are plotted for
truncation at $l=1,4,16$ (left to right). All demonstrate unphysical
artefacts from the truncation; convergence to a smooth solution is
only achieved at $l=32$, shown in the bottom-left panel of Fig. \ref{mix_field}.}
\end{minipage}
\end{center}
\end{figure*}

Solution of the magnetic-equilibrium equations requires a
multipolar decomposition of the exact vector equations into an
infinite series of scalar equations in terms of different
multipoles. Clearly one cannot in practice solve this infinite system,
and must truncate the multipolar expansion at some value of the angular
index. For semi-analytic models (e.g. \citet{ciolfi10}) it is only
practicable to retain a few multipoles at best; in our numerical
study we have the luxury of producing equilibria including the
contributions of higher multipoles with little extra computational
time.

In solving for a large-scale, global magnetic-field equilibrium, it is
natural to expect the solution to be dominated by low-multipole
contributions -- but it is not clear how many multipoles should be
retained for a faithful approximation to the exact infinite-multipole
result. We check this in Fig. \ref{multipole_plot}, for a linked
poloidal-toroidal field whose toroidal component is very intense and
localised in the outer equatorial region. We show only the poloidal
component -- strength and magnitude -- since the toroidal component
looks similar in each case. We find a major difference in the
magnetic-field structures coming from truncating at low and high multipoles, and even
between the $l=16$ and $32$ models; truncation for $l>32$, on the
other hand, makes little difference. This is also seen in the ratio
$\mathcal{E}_{\rm mag}^{\rm tor}/\mathcal{E}_{\rm mag}$, which is
$2.4\%,3.4\%,8.2\%,8.5\%$ respectively for $l=1,4,16,32$.

We have undertaken similar comparisons for other cases. They are not
plotted for reasons of brevity, but we find that high multipoles are
similarly important in any magnetic-field model with significant
stellar distortion (either from the magnetic field or rotation), but
less so for almost-spherical poloidal-toroidal models without strong toroidal
components. Only in this latter case (dominantly poloidal fields) is
it safer to terminate at low multipoles.

In almost-spherical stars
without extremely strong toroidal field components, the solution is
seen to be dominated by the dipole component. However, for very
intense toroidal components and/or significant stellar distortion, we
see that a large number of multipoles must be summed before artefacts
of the truncation cease to be visible.

\section{Discussion}
\label{discussion}

The primary motivation for undertaking this work was to study
differences between the magnetic fields of young and mature NSs. They
have turned out to be very similar, a result that raises more questions
than it answers. In closing, it is therefore natural to discuss the implications
of this result, and how realistic and general our
results are.

\subsection{Comparing cold and hot models}

There were reasons to anticipate differences between cold and hot
models. The strong thermal pressure -- accounting for a substantial
fraction of the star's total pressure for our hottest models --
represents a new piece of physics compared with a cold star. At
  the same time, we have not accounted for any kind of buoyancy force,
  even though they may play a role in a real proto-NS's equilibrium; see section
\ref{eqm_EOS}. The field is again governed by the Grad-Shafranov equation
for barotropic fluids, whose solutions are only weakly affected by
differences in the star's pressure/density distribution. The
core's thermal pressure $P_{\rm th}\propto\rho^{5/3}$, which is not
significantly different from the adiabatic index of $2-3$ for the
zero-$T$ core
pressure. We believe these two effects -- the similar
pressure distribution and the lack of
buoyancy force -- are the key reasons why the magnetic fields of hot
equilibria are so similar to their cold counterparts. It also suggests
that a relativistic version of our Newtonian equilibrium model -- essentially
amounting to changing the gravitational potential -- would give similar results.

\subsection{Relative strengths of poloidal and toroidal components}

All of our new magnetic-field configurations for hot NSs are -- like their
cold predecessors -- energetically dominated by
the poloidal field component (the only exception being \emph{purely}
toroidal fields -- but these are unstable and, having no exterior
component, would not be directly observable). This is in conflict with a number of
other pieces of work that rely on a NS's magnetic field being
dominated by its toroidal component. Typical supporting evidence invoked for
dominantly-toroidal fields is the work of \citet{braith_09} and
\citet{akgun13}, but we argue that the strong buoyancy forces required
to support these equilibria may not exist in the proto-NS phase (and
perhaps not at later stages either). To our knowledge the only
barotropic NS model with a dominantly toroidal field is that presented
in \citet{ciolfi_rezz}. With a careful choice of the magnetic
functions $f(u)$ and $M(u)$ (see section \ref{func_choice}), they were
able to control the magnetic-field structure and produce a much wider
range of poloidal- and toroidal-component strengths; see also
\citet{fuji_eri} for a physical interpretation of this choice.

We have also experimented with a range of different choices for
$f(u),M(u)$, including those of \citet{ciolfi_rezz}, but all of our
resulting equilibria resemble those of Fig. \ref{mix_field} instead of
ever having dominant toroidal components. There are two significant
differences in our approach: firstly, our study involves numerical
solution for self-consistent equilibria rather than an essentially
analytic approach; secondly, that we retain a
far higher number of multipoles in our solution (\citet{ciolfi_rezz}
only allowed for a dipole, $l=1$, field component). During our iterations
we observe that even if we start with a larger region of toroidal
field, as engineered by a
careful choice of $f(u),M(u)$, this shrinks rapidly before the
iterative method converges. How can we explain this disagreement? One
possible scenario is that there is more than one
branch of solutions to the Grad-Shafranov equation, and that our code
`picks' only a particular poloidal-dominated one. Another -- and we believe more likely --
possibility is that the results of \citet{ciolfi_rezz} are only
approximate equilibria, resulting from truncating at the dipole
component and not considering the backreaction of the field -- and 
that true self-consistent equilibrium models all resemble those we
present in this work. We have already seen the dangers of truncating at low
multipoles: in section \ref{multipoles}, it was shown to cause serious errors in
the resulting field configurations, including in the ratio
$\mathcal{E}_{\rm mag}^{\rm tor}/\mathcal{E}_{\rm mag}$.

Further evidence for the universality of our poloidal-dominated
equilibria is that at least 
two other independent numerical studies have also used the
prescription of \citet{ciolfi_rezz} \emph{without} managing to obtain
toroidal-dominated equilibria (\citet{bucc15}; Armaza, private
communication). The common feature of all three numerical codes seems
to be the retention of high multipoles in the solutions; 
\citet{fuji_eri} also pointed out the likely
importance of higher multipoles in their analysis of this
problem. Finally, we note that non-linear evolutions of
of an initially unstable magnetic field tend to show saturation to a state
(albeit a dynamic one, not a strict equilibrium) with
$\mathcal{E}_{\rm mag}^{\rm tor}/\mathcal{E}_{\rm mag}\lesssim 25\%$
\citep{lasky11,ciolfi11,sur}.

\subsection{Stability and rearrangement of the field}

We have argued that the magnetic field in a late proto-NS can be
reasonably described as an equilibrium, and that such an equilibrium configuration
appears to be poloidal-dominated in all cases. This has several
implications. Firstly and most seriously, it has been argued that a stable
magnetic equilibrium needs both poloidal and toroidal components, with
at least comparable energies \citep{tayler80} -- which would imply there are no
astrophysically relevant equilibria for proto-NSs at
all. Qualitatively similar magnetic fields in zero-temperature models
have been shown to be unstable \citep{LJ12}, with the instability for
poloidal-dominated fields developing in the region of closed field
lines (where the toroidal component is also present). A glimmer of hope for the
models presented here is that the temperature gradient may have a
stablising effect; in addition, although the toroidal component is not
energetically dominant, it can be locally comparable in strength with
the poloidal one in the most unstable region of the star.

A second implication of our results is that
a number of scenarios relying on a newborn NS having a strong toroidal
field may be irrelevant, if no such field configuration exists. It is quite conceivable that differential
rotation drives a strong amplification of the toroidal field component
shortly after birth, but once this driving force ceases the field must
rearrange into a state like our models. This suggests that at this
early stage the magnetic field may shed a considerable amount of
energy in its attempt to become an equilibrium state -- an event
likely to be powerful enough for detection. Furthermore, we have found
that very rapidly-rotating stars can support stronger toroidal fields
than non-rotating ones; this hints at another possible source of
energy release from magnetic-field rearrangement on the star's
spindown timescale.

All our conclusions apply to relatively strong magnetic fields; see
section \ref{quasi_eq}. If instead the birth field is weak, so that the characteristic time for
rearrangement is longer than the cooling timescale, we
anticipate that it may avoid substantial rearrangement. Weaker NS
magnetic fields could then have qualitatively different structures
from stronger ones.

\subsection{Lower break-up velocity}

We find that a very hot proto-NS reaches break-up (Keplerian) velocity
at a far lower rotational frequency than a cold model: by a factor
of about a third. Our piecewise-polytropic treatment of the cold
equation of state leads to a value of $\nu_K=960$ Hz for a
$1.4\mathcal{M}_\odot$ cold star, in excellent agreement with 
the value $\nu_K=970$ Hz resulting from the approximate formula
in \citet{haensel09} (within the range of accuracy of this approximation), 
but this drops below $\nu_K=700$ Hz for
the hottest models, which can be explained by the significantly larger
radii of these stars. Since the star is born hot, it is this latter,
smaller value of $\nu_K$ that sets the effective limiting rotation
rate in the star's early life. Note that for all plausible field strengths
($B\lesssim 10^{17}$ G), the magnetic field has no effect on the value
of $\nu_K$ \citep{LJ09}. This low value of $\nu_K$ may make it more difficult to
realise various interesting scenarios: gravitational waves
from unstable $r$- or $f$-modes in rapidly-rotating newborn NSs, or millisecond
magnetars and their associated electromagnetic/gravitational radiation. Our limiting
rotation rate is however not relevant for explaining the puzzle of
rotation rates of old, recycled NSs having an upper limit well below
$\nu_K$, since for this scenario the cold value is relevant.

\subsection{The future of the proto-NS's magnetic field}

The majority of observed NSs have strong magnetic fields with large-scale
structure. They have no obvious mechanism for regeneration of magnetic
flux, indicating that the field remaining at the end of the proto-NS
phase is not substantially dissipated over thousands of years. We have
argued, however, that instabilities may plague our models -- and such
instabilities involve widespread disruption to the magnetic field and
turbulent fluid motions, which are likely to cause a major loss of
magnetic energy.

The resolution to this contradiction could be an additional piece of
proto-NS physics -- for example, if differential rotation persists
into this late stage and allows for a stronger toroidal field than in
our models -- or the advent of a new stabilising mechanism as the star
cools. A day into its life, with a temperature not much above $10^9$
K \citep{gnedin}, a NS will have started developing two such
candidate mechanisms: a modest but growing region of
superconducting protons in its core, and solidification of its
envelope into a crust, starting from the boundary with the core and
slowly moving outwards. Both may inhibit magnetic
instabilities: the former by changing the local structure and dynamics
of the field, and the latter by providing an elastic force to resist
unstable motion.

Two factors may assist the crust in stabilising the stellar magnetic
field. Firstly, although the toroidal-field component
might not itself stabilise the poloidal-dominated field, it can help
indirectly by pushing the unstable closed-field-line region
outwards into the crust (see section \ref{mag_struct}). Secondly, a
strong magnetic field induces the formation of an extended inner crust
region \citep{fang}, which could be as much as $\sim 1$ km in a
magnetar, thus increasing the likelihood of the closed-field-line
region coinciding with the crust \citep{sengo}.

\subsection{Outlook}

The study of NS magnetic fields has reached a juncture, where
quantitative models tend to include only very simple physics, and
where consideration of more realistic physics is often speculative and
qualitative. Quantitative studies of the birth phase of NS magnetic fields are likely to be crucial to
improving this situation: the dynamo processes generating magnetic
flux straight after birth, the immediate post-dynamo phase in which
the field should presumably relax into an equilibrium, and the later
formation of a solid crust and superconducting regions in the core. We
have tried, in this work, to take a first step in that direction.


\section*{Acknowledgements}

SKL acknowledges support from the European Union's Horizon 2020
research and innovation programme under the Marie Sk\l{}odowska-Curie
grant agreement No. 665778, via fellowship UMO-2016/21/P/ST9/03689 
of the National Science Centre (NCN), Poland; MF was supported by NCN
grant 2017/26/D/ST9/00591. We thank Crist\'obal Armaza
for helpful correspondence, the referees for useful criticism, and the
PHAROS COST Action (CA16214) for partial support.

\section*{Data availability}

The specific data underlying this article, and additional data for
other stellar models, will be made available upon reasonable request.

\bibliographystyle{mnras}


\bibliography{references}

\label{lastpage}

\end{document}